\documentclass[aps,prf,twocolumn,groupedaddress,longbibliography]{revtex4-1}
\usepackage{graphicx}
\usepackage{dcolumn}
\usepackage{bm}
\usepackage{longtable}
\usepackage{float}
\usepackage{tikz}
\usepackage{amsmath}
\usepackage{amssymb}
\usepackage{breqn}
\usepackage{etoolbox}

\newrobustcmd*{\myVtriangle}[2]{\tikz{\filldraw[draw=#1,fill=#2] (0cm,0.2cm) --
(0.2cm,0.2cm) -- (0.1cm,0cm) -- (0cm,0.2cm);}}

\newrobustcmd*{\mythickVtriangle}[2]{\tikz{\filldraw[line width=0.3mm,draw=#1,fill=#2] (0cm,0.2cm) --
(0.2cm,0.2cm) -- (0.1cm,0cm) -- (0cm,0.2cm);}}

\newrobustcmd*{\mytriangle}[2]{\tikz{\filldraw[draw=#1,fill=#2] (0.0cm,0.0cm) --
(0.2cm,0cm) -- (0.1cm,0.2cm) -- (0cm,0cm);}}

\newrobustcmd*{\mysquare}[2]{\tikz{\draw[draw=#1,fill=#2] (0cm,0cm)
rectangle (0.2cm,0.2cm)}}

\newrobustcmd*{\mythicktriangle}[2]{\tikz{\filldraw[line width=0.3mm,draw=#1,fill=#2] (0.0cm,0cm) --
(0.2cm,0cm) -- (0.1cm,0.2cm) -- (0.0cm,0cm);}}

\newrobustcmd*{\mythicksquare}[2]{\tikz{\draw[line width=0.3mm,draw=#1,fill=#2] (0cm,0cm)
rectangle (0.2cm,0.2cm)}}

\newrobustcmd*{\mybarredtriangle}[2]{\tikz{\draw[draw=#1,fill=#2] (0,0) --
(0.2cm,0) -- (0.1cm,0.2cm) -- (0cm,0cm); \draw[draw=#1] (-0.1cm, 0.07cm) -- (0.3cm, 0.07cm)}}

\newrobustcmd*{\mythickbarredtriangle}[2]{\tikz{\draw[line width=0.3mm,draw=#1,fill=#2] (0,0) --
(0.2cm,0) -- (0.1cm,0.2cm) -- (0cm,0cm); \draw[draw=#1] (-0.1cm, 0.07cm) -- (0.3cm, 0.07cm)}}

\newrobustcmd*{\mybarredsquare}[2]{\tikz{\draw[draw=#1,fill=#2] (0,0)
rectangle (0.2cm,0.2cm); \draw[draw=#1] (-0.1cm, 0.1cm) -- (0.3cm, 0.1cm)}}

\newrobustcmd*{\mythickbarredsquare}[2]{\tikz{\draw[line width=0.3mm,draw=#1,fill=#2] (0,0)
rectangle (0.2cm,0.2cm); \draw[draw=#1] (-0.1cm, 0.1cm) -- (0.3cm, 0.1cm)}}

\newrobustcmd*{\mybarredcircle}[2]{\tikz{\draw[draw=#1,fill=#2] (0,0)
circle (0.1cm); \draw[draw=#1] (-0.2cm, 0.0cm) -- (0.2cm, 0.0cm)}}

\newrobustcmd*{\mythickbarredcircle}[2]{\tikz{\draw[line width=0.3mm,draw=#1,fill=#2] (0,0)
circle (0.1cm); \draw[draw=#1] (-0.2cm, 0.0cm) -- (0.2cm, 0.0cm)}}

\newrobustcmd*{\mydashedline}[1]{\tikz{\draw[draw=#1] (-0.2cm, 0.2cm) -- (-0.1cm, 0.2cm); \draw[draw=#1] (-0.0cm, 0.2cm) -- (0.1cm, 0.2cm)}}

\newrobustcmd*{\mythickcross}[1]{\tikz{\draw[line width=0.3mm,draw=#1] (0,0) --
(0.2cm,0); \draw[line width=0.3mm,draw=#1] (0.1cm,-0.1cm) -- (0.1cm,0.1cm);}}

\newrobustcmd*{\mybarredcross}[1]{\tikz{\draw[line width=0.3mm,draw=#1] (0,0) --
(0.2cm,0); \draw[line width=0.3mm,draw=#1] (0.1cm,-0.1cm) -- (0.1cm,0.1cm); \draw[draw=#1] (-0.1cm,0) -- (0.3cm,0);}}

\newrobustcmd*{\myline}[1]{\tikz{\draw[draw=#1] (-0.15cm, 0.1cm) -- (0.15cm, 0.1cm);\draw[line width=0.3mm,draw=#1] (-0.0cm, 0.0cm);}}

\newrobustcmd*{\mythickline}[1]{\tikz{\draw[line width=0.3mm,draw=#1] (-0.15cm, 0.1cm) -- (0.15cm, 0.1cm);\draw[line width=0.3mm,draw=#1] (-0.0cm, 0.0cm);}}

\newrobustcmd*{\mythickdashedline}[1]{\tikz{\draw[line width=0.3mm,draw=#1] (-0.2, 0.1cm) -- (-0.1cm, 0.1cm); \draw[line width=0.3mm,draw=#1] (-0.0cm, 0.1cm) -- (0.1cm, 0.1cm); \draw[line width=0.3mm,draw=#1] (-0.0cm, 0.0cm);}}

\newrobustcmd*{\mythickdasheddottedline}[1]{\tikz{\draw[line width=0.3mm,draw=#1] (-0.22, 0.1cm) -- (-0.13cm, 0.1cm); \draw[line width=0.3mm,draw=#1] (-0.085cm, 0.1cm) -- (-0.055cm, 0.1cm); \draw[line width=0.3mm,draw=#1] (-0.01cm, 0.1cm) -- (0.08cm, 0.1cm); \draw[line width=0.3mm,draw=#1] (-0.0cm, 0.0cm);}}

\newrobustcmd*{\mycircle}[2]{\tikz{\draw[draw=#1,fill=#2] (0,0)
circle (0.1cm);}}

\newrobustcmd*{\mythickcircle}[2]{\tikz{\draw[line width=0.3mm,draw=#1,fill=#2] (0,0)
circle (0.1cm);}}

\newrobustcmd*{\mydot}[1]{\tikz{\draw[line width=0.3mm,draw=#1] (0,0)
circle (0.025cm);}}

\usepackage{epstopdf, epsfig}

\newcommand{\bxi}{\boldsymbol{\xi}}
\newcommand{\bg}{\boldsymbol{\Gamma}}
\newcommand{\bgsv}{\boldsymbol{g}}
\newcommand{\bx}{\boldsymbol{x}}
\newcommand{\bk}{\boldsymbol{\kappa}}
\newcommand{\bphi}{\boldsymbol{\phi}}

\newcommand{\norm}[1]{\left\lVert#1\right\rVert}

\begin{document}

\title{Lyapunov spectrum of forced homogeneous isotropic turbulent flows}

\author{Malik Hassanaly}
 \email{malik.hassanaly@gmail.com}
\author{Venkat Raman}%
 \email{ramanvr@umich.edu}
\affiliation{%
Department of Aerospace Engineering \\ University of Michigan \\ Ann Arbor MI 48109 United States
}%



\begin{abstract}
In order to better understand deviations from equilibrium in turbulent flows, it is meaningful to characterize the dynamics rather than the statistics of turbulence. To this end, the Lyapunov theory provides a useful description of turbulence through the study of the perturbation dynamics. In this work, the Lyapunov spectrum of forced homogeneous isotropic turbulent flows is computed. Using the Lyapunov exponents of a flow at different Reynolds numbers, the scaling of the dimension of the chaotic attractor for a three-dimensional homogeneous isotropic flow (HIT) is obtained for the first time through direct computation. The obtained Gram-Schmidt vectors (GSV) are analyzed. For the range of conditions studied, it was found that the chaotic response of the flow coincides with regions of large velocity gradients at lower Reynolds numbers and enstrophy at higher Reynolds numbers, but does not coincide with regions of large kinetic energy. Further, the response of the flow to perturbations is more and more localized as the Reynolds number increases. Finally, the energy spectrum of the GSV is computed and is shown to be almost insensitive to the Lyapunov index. 
\end{abstract}
\maketitle

\section{Introduction}
\label{sec:intro}
The dynamical systems (DS) approach to representing chaotic flows provides insights that are generally different from statistical approaches \cite{ruelle1971nature,constantin1985determining,keefe,mohan2017scaling,kalnayBook}. With the emergence of data-driven modeling, there has been a renewed focus on the DS viewpoint of turbulence~\cite{lu2017data}. From a physics standpoint, the statistical view provides a natural pathway to modeling in the form of averaged equations, which itself may take different forms \cite{popebook,smagor,moser-optimal,pope2010self}. However, such statistical approaches inherently model or represent only the average dynamics of turbulent flow \cite{adrian88,adrian90,moser-optimal,pope2010self}. If the deviations from the average behavior are of interest, then the DS approach might provide a better starting point \cite{kalnayBook,buizza}. In recent years, there has been growing interest in the prediction of extreme events \cite{peherstorfer2018survey,farazmand2018extreme,raman2018emerging,malik_msc10}, defined as anomalous excursions of the non-linear DS from an expected average path. Hassanaly and Raman describe multiple causal mechanisms for such extreme events in the context of turbulent reacting flows \cite{raman2018emerging,hassanaly2019computational,malik_msc10}. The description of such extreme events is shown to be linked to the structure of the phase space, in particular, the strange attractor that defines the chaotic system \cite{tailleur,malik_msc10}. Characterizing such attractors is often carried out using Lyapunov theory \cite{oseledets1968multiplicative, benettin1,benettin2,shimada1979numerical,ruelle,ruelle1979ergodic} but can also be computed using other techniques \cite{budanur2017relative,kevlahan2007scaling}. The focus of this work is on the use of Lyapunov theory for a canonical turbulent flow configuration, namely, homogeneous isotropic turbulence (HIT).

In the Lyapunov theory, the response of the turbulent flow fields to perturbations is analyzed. From the DS perspective, the turbulent flow is represented using a set of ordinary differential equations that are obtained by an appropriate discretization of the governing equations. The Lyapunov analysis provides the growth rate of specifically-aligned perturbations to the state of the system. The growth rate itself is expressed as the exponential of Lyapunov exponent (LE), while the corresponding perturbation is termed a Lyapunov vector (LV) (more formal definitions provided in Sec.~\ref{sec:numerics_lyap}). For a $\mathcal{D}$-dimensional system, $\mathcal{D}$ such pairs of Lyapunov exponents and vectors can be defined. Several algorithms have been used to extract the LEs and LVs~\cite{benettin2,shimada1979numerical,ginelli}, which may be thought of as equivalent to eigenvalue/eigenvector pairs of the Oseledets matrix, defined using the backward dynamics \cite{ginelli2013covariant,malik_thesis}. In general, the LEs are ordered from largest and smallest, and for any chaotic system, will possess positive and negative values. The knowledge of these quantities can lead to a description of the process by which instabilities occur in a turbulent flow. Using LVs, it was found, for example, that the chaotic behavior in a Taylor-Couette flow occurred due to a Kelvin-Helmholtz instability~\cite{vastano_moser}. Similarly, streamwise vortices were found to be at the origin of streak regeneration in Couette turbulence~\cite{inubushi2015regeneration}. More recently, the importance of walls in the generation of instabilities in Rayleigh-B\'{e}nard convection was deduced from the Lyapunov analysis~\cite{xu2018chaotic}. In the context of turbulent combustion, the Lyapunov analysis allowed to find that the chaotic behavior of a turbulent jet flame was mostly due to intermittent ignition/extinction~\cite{hassanaly2019ensemble}. The Lyapunov spectrum (LS) also carries important information about the long-time behavior of a dynamical system, or more specifically, the attractor. The LE can provide an estimate of the dimension of the attractor~\cite{kaplan1979chaotic,keefe}, while the LV can inform about its geometry (hyperbolicity for example as in \cite{xu2016covariant} and \cite{inubushi2012covariant}). In terms of quantitative predictions, the knowledge of the first LE (which is the largest LE) provides an estimate of the decorrelation time. In other words, the inverse of the LE is a time-scale that determines the time horizon of predictions with a prescribed level of uncertainty \cite{lorenz,kalnayBook}. In the context of HIT, which is the main focus here, the value of the scaling of the first LE has been obtained in previous work~\cite{mohan2017scaling}. 

The focus here is on the set of Lyapunov exponents and vectors for HIT. While there have been many studies on extracting the spectra for other canonical flows, such as the Kuramoto-Sivashinsky (KS) system~\cite{yang2009hyperbolicity} and the turbulent channel flow~\cite{keefe}, the periodic turbulent flow in HIT has not been explored. Even so, the focus of many of these studies has been on the Lyapunov exponents. Since the computation of each exponent requires one additional forward run of the DS, obtaining a large number of such exponents can quickly become computationally intractable, depending on the complexity of the DS equations. When the DS is not high-dimensional, then the entire spectrum can typically be computed \cite{karimi2010extensive,eckmann2005lyapunov}. More specifically, all the positive exponents of the system are directly computed, while retaining some of the negative exponents. Since the positive and the first few negative exponents determine the structure of the attractor, and the most negative exponents track the dissipative behavior, such extensive Lyapunov computations provide detailed information about the chaotic dynamics of the system.

With this background, the goal of this work is to compute and study the Lyapunov spectrum for HIT. Section~\ref{sec:numerics} treats the numerical approach used for the DS and the computation of the LS. Section~\ref{sec:results} covers the results obtained by analyzing both the LE and the LV. In particular, the effect of different Reynolds number on the dimension of the chaotic attractor (Sec.~\ref{sec:scaling}) and of the forcing schemes on the LE (Appendix~\ref{sec:app_forcingLE}) are discussed. The response of the flow field to perturbation is analyzed in Sec.~\ref{sec:lv} by examining the backward Lyapunov vectors, or Gram-Schmidt vectors (GSV). Their structure, as well as their dependence on the Reynolds number, is investigated. The findings are summarized and discussed in Sec.~\ref{sec:conclusion}.


\section{Configuration and computational approach}
\label{sec:numerics}
The simulation configuration and the numerical procedure for extracting the Lyapunov exponents and vectors are briefly described below. For a full description, including the numerical convergence details, the reader is referred to \cite{hassanaly2019numerical}. 

\subsection{Flow configuration}
Forced homogeneous isotropic turbulence in the low Mach number incompressible flow regime is simulated in a $2\pi$-triply periodic box. A constant density of $\rho=1$~kg.m$^{-3}$ and kinematic viscosity of $\nu=0.05$~m$^2$.s$^{-1}$ are used throughout the paper. The governing equations of the DS are written as
\begin{equation}
 \frac{\partial u_i}{\partial t} + \frac{\partial u_i u_k}{\partial x_k} = -\frac{1}{\rho}\frac{\partial p}{\partial x_i} + \nu\frac{\partial}{\partial x_k} \frac{\partial u_i}{\partial x_k} + f_i,
 \label{eq:ns}
\end{equation}
\begin{equation}
 \frac{\partial u_i}{\partial x_i} = 0,
 \label{eq:cont}
\end{equation}
where $u_i$ is the velocity component, $f_i$ is a forcing term in the $i$-th direction that maintains the turbulence level, and $p$ is the fluid pressure. In absence of forcing, the turbulent kinetic energy in the domain tends to zero due to viscous dissipation. Therefore, the forcing term ensures statistical stationarity, which is necessary for evaluating the Lyapunov spectrum. The impact of the forcing term will be studied in detail in Appendix~\ref{sec:app_forcingLE}. For this reason, the exact form of the forcing term is discussed later.

\subsection{Numerical details}
\label{sec:numerics_lyap}

In a dynamical system context, the discretized governing equations (Eq.~\ref{eq:ns} and \ref{eq:cont}) are written as a set of ordinary differential equations that takes the form 
\begin{equation}
    \frac{d\boldsymbol{\xi}}{dt} = \mathcal{F}(\boldsymbol{\xi});~~\boldsymbol{\xi}(t=0)=\boldsymbol{\xi}^0,
\end{equation}
where $\boldsymbol{\xi}$ is a vector of all the variables, $\mathcal{F}$ is the discretized form of Eq.~\ref{eq:ns}, and $\boldsymbol{\xi}^0$ are the initial conditions. The vector $\boldsymbol{\xi}$ is also called a state vector as it describes the state of the system. Here, because the pressure can be readily obtained from the velocity using the incompressibility condition, the state vector is composed of all the variables describing only the velocity field at a certain instant. For a fluid problem in three spatial dimensions, the state vector is of dimension $\mathcal{D} = 3 \times N^3$, where $N$ is the number of grid points or Fourier modes in one direction. Several resolutions are used investigated ranging from $32^3$ to $128^3$ modes/grid points, resulting in  $\mathcal{D} \sim 10^5 -  10^7$. The present problem is high-dimensional, which introduces a computational hurdle in the study of its chaotic behavior. 

The Lyapunov analysis consists of studying the evolution of perturbations $\delta \boldsymbol{\xi}$ applied to the system in this high-dimensional space. These perturbations evolve according to the following equation:
\begin{equation}
\frac{d\delta\boldsymbol{\xi}}{dt} = \frac{\partial \mathcal{F}}{\partial \boldsymbol{\xi}}\delta \boldsymbol{\xi} = \mathcal{J}(\boldsymbol{\xi})\delta \boldsymbol{\xi} ;~~\delta{\boldsymbol{\xi}}(t=0) = \delta{\boldsymbol{\xi}}^0,   
\label{eq:pertEq}
\end{equation}
where $\mathcal{J}$ is the linearized evolution operator of the perturbation (the Jacobian of the dynamics). As the perturbations advance in time, their norm is subject to an exponential expansion or contraction rate, which depends on the unperturbed state vector and the direction of the perturbation. For every state vector, the rate of variation of the perturbation norm can be decomposed into $\mathcal{D}$ individual rates associated with particular initial perturbation direction. In the long-time limit, it was shown by Oseledets \cite{oseledets1968multiplicative} that in an ergodic system, there exist limits for these expansion rates, which are called the Lyapunov exponents. The very existence of this limit is what allows to adopt a global point of view on the dynamics of the DS, as opposed to examining each point in phase-space. The computation of the LE and associated LV requires distinguishing between different perturbation directions that expand at different rates. Here, the algorithm of Benettin et al.~\cite{benettin2} is used. This algorithm computes the first $m$ LE, where $m \leq \mathcal{D}$, by evolving $m$ copies of Eq.~\ref{eq:pertEq}. The perturbations are periodically orthogonalized to identify directions that expand at different rates. In the end, the set of LEs obtained is noted as: 
\[ \boldsymbol{\lambda} \equiv \{ \lambda_1, ..., \lambda_m\} \]
In the rest of the paper, $\boldsymbol{\lambda}$ is referred to as the \textit{Lyapunov spectrum} (LS). As shown in previous work~\cite{hassanaly2019numerical}, the perturbation can initially grow due to the numerical effects rather than due to the dynamics of the system. It is useful in that case to wait for a few timesteps before recording the norm of the perturbations. At best, this allows eliminating the spurious perturbation growth. At worst, it recovers the same perturbation growth rate. Note that the waiting period only affects the computation of the LE. Here, this procedure is used for all the simulations with a waiting time of $0.2$ s (10 to 150 timesteps depending on the resolution) and an averaging time of the perturbations growth of one eddy turnover time s (300 to 1300 timestep depending on the resolution). The initial size of perturbations is chosen such that they remain small compared to the baseline flow during the growth phase (See Appendix~\ref{app:perturbationSize}). The set of LVs obtained (GSV) from the procedure are required to be orthogonal (here according to the Euclidean scalar product) to each other and were shown to be independent of the initial perturbation \cite{ershov1998concept}. The GSVs are orthogonal by definition and depend on the scalar product chosen. A less ambiguous set of vectors are the covariant Lyapunov vectors (CLVs) which do not require the definition of a scalar product. The CLVs are defined as the vectors that evolve with the tangent dynamics \cite{ruelle1979ergodic,ginelli2013covariant,malik_thesis}. Nevertheless, since the GSV converge to the backward Lyapunov vectors \cite{ershov1998concept} they describe the response of the system to perturbations that expand or contract and thereby provide interesting information about the DS. Since the CLV can also be expressed as a linear combination of the GSV, studying the GSV allows drawing conclusions about the tangent space to the attractor. This aspect is further discussed in Appendix~\ref{sec:app_clv}. In the rest of the paper, LV always refers to GSV unless specified otherwise.

Instead of solving Eq.~\ref{eq:pertEq}, the growth in perturbations is obtained by solving a baseline simulation and perturbed simulations, with the difference in the solution vectors at a future time providing the perturbation at that time. This procedure circumvents the need to compute the Jacobian matrix, which may become computationally expensive, especially for large systems with complex evolution equations. Further, assembling the Oseledet matrix (see \cite{ginelli2013covariant} for a precise definition) would lead to large numerical errors. To obtain the $m$ LEs/LVs, $m$ perturbed simulations of a baseline calculation are evolved. The approach of Benettin \cite{benettin1,benettin2} is then used to obtain the LEs and LVs. Previously, this numerical procedure has been verified for a number of canonical flows using a low-Mach number flow solver \cite{hassanaly2019numerical}. The low-Mach solver is based on the NGA code \cite{desjardins-jcp}, and the underlying numerical algorithms have been tested in \cite{malik-openfoampaper,heye_proc}. This code has been validated over a large range of turbulent flows in the past, including for HIT cases \cite{carroll2013proposed,palmore2018technique}. The code solves Eq.~\ref{eq:ns} in physical space using a staggered arrangement of variables in space and time. Second-order space and time discretization are used for integrating Eq.~\ref{eq:ns}. The incompressibility condition is enforced through a Poisson equation at the end of each timestep: the pressure is computed such that the velocity is a divergence-free field at the end of every timestep. Convergence results for the computations of the Lyapunov exponents using this code are available in Ref.~\cite{hassanaly2019numerical}.

In the present work, the incompressible governing equations (Eq.~\ref{eq:ns}) are solved using a Fourier spectral transformation \cite{orszag1972numerical,moin1998direct}, by decomposing the primitive variables into Fourier modes:
\[
u_i  = \sum_{\bk} \widehat{u_i}(\bk,t) e^{j \bk\cdot \bx}
\]
where $\bk$ is the wave vector and $\bx$ is the vector of the location in physical space. The Galerkin projection of Eq.~\ref{eq:ns} is obtained for all the Fourier modes. A pseudo-spectral method with dealiasing is used for the non-linear term and is integrated in time using a second-order Runge-Kutta scheme. The viscous term is integrated analytically. The timestep is chosen so as to maintain a Courant number of $0.5$ for all the simulations. The ratio of the Kolmogorov length scale to grid size is maintained close to unity. Due to round-off numerical errors, it was found that there was a growth of instability in the simulations that contaminated the Lyapunov evaluations. For this reason, the continuity is explicitly forced using a correction procedure at each time step (See \cite{malik_thesis} for details). 

As a validation of the implementation, the LS of the same case as the one obtained in \cite{hassanaly2019numerical} which used a spatial discretization is computed this time with the spectral code. A linear forcing \cite{lundgren_linearforcing,Rosales:2005fy} with a coefficient $A=0.1$ is used and Eq.~\ref{eq:ns} takes the form 
\begin{equation}
 \frac{\partial u_i}{\partial t} + \frac{\partial u_i u_k}{\partial x_k} = -\frac{1}{\rho}\frac{\partial p}{\partial x_i} + \nu\frac{\partial}{\partial x_k} \frac{\partial u_i}{\partial x_k} + A u_i.
\label{eq:ns_lin}
\end{equation} 

\begin{figure}[hb]
\begin{center}
\includegraphics[width=0.4\textwidth,trim={0.1cm 0.1cm 0.1cm 0.1cm},clip]{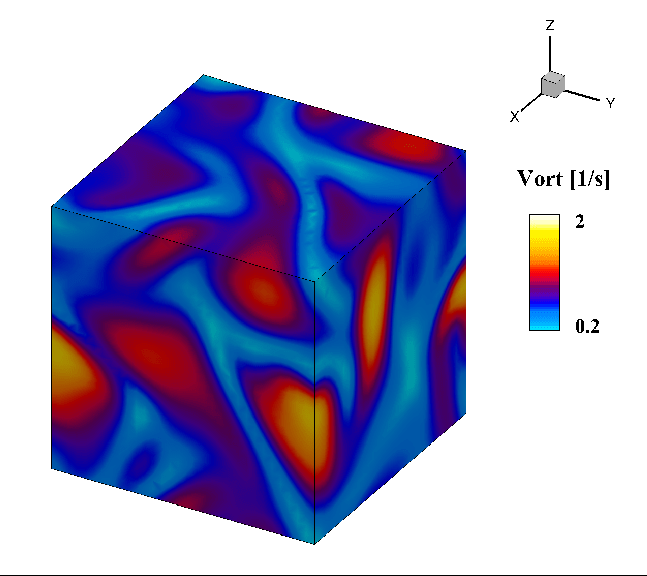}
\includegraphics[width=0.4\textwidth,trim={0cm 0cm 0cm 0cm},clip]{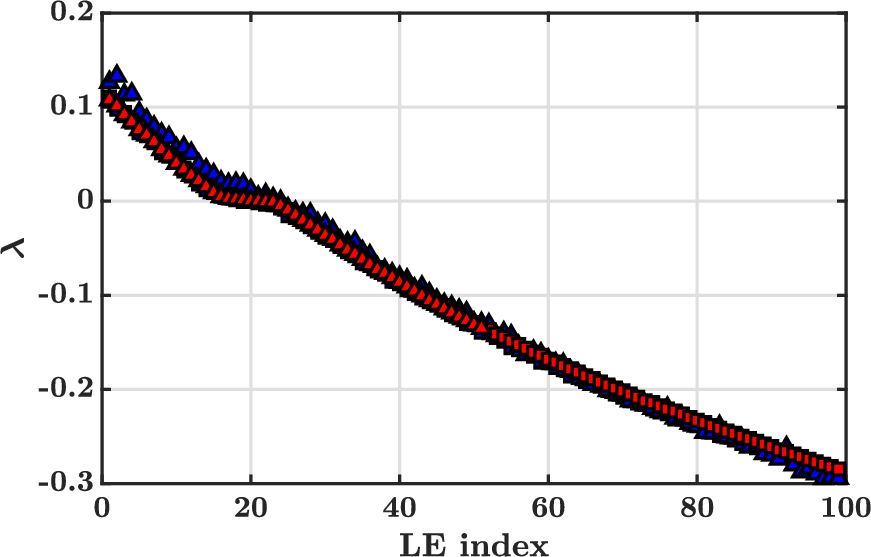}
\caption{Top: contour of vorticity magnitude for the case solved with the spectral code using $32^3$ Fourier modes. Bottom: first 100 LE obtained with: the spectral code with $32^3$ Fourier modes  (\mythicksquare{black}{blue}) and $64^3$ Fourier modes (\mythicktriangle{black}{blue}); the physical space code with $32^3$ grid points (\mythicksquare{black}{red}) and $64^3$ grid points (\mythicktriangle{black}{red}, first 50 exponents only).}
\label{fig:LEvalid}
\end{center}
\end{figure}

In order to verify the spectral code, four different simulations are run: a) $32^3$ and $64^3$ simulations with the low-Mach number solver (same as the one presented in Ref.~\cite{hassanaly2019numerical}, b) $32^3$ and $64^3$ Fourier mode simulations with the spectral solver. Figure~\ref{fig:LEvalid} (top) shows a snapshot of the vorticity magnitude, demonstrating the presence of well-defined vortical structures in the relatively low Reynolds number flow. For all $32^3$ calculations, 100 LEs were computed, while 50 LEs were computed for the $64^3$ spectral simulation. The Lyapunov spectra from all of these calculations are shown in Fig.~\ref{fig:LEvalid} (bottom). The good agreement between all four cases indicates that the dynamics of the flow were sufficiently resolved to capture the spectrum. It also shows that both the low-Mach spatial solver and the spectral solver capture similar dynamics of the flow field. 

\section{Results}
\label{sec:results}

In this section, the LEs are used to directly estimate the dimension of the attractor of the flow field. Using different Reynolds numbers, an estimate of the scaling of the attractor dimension is also obtained. The structure of the LVs is also described; their localization in physical space is characterized as well as their correlation with classical turbulent flow quantities.

\subsection{LE spectrum}

One of the main uses of the Lyapunov study is the determination of the dimension of the attractor. For this purpose, a set of Lyapunov exponents needs to be computed. In order to obtain an accurate estimate, it is necessary to determine at least all of the positive exponents and some of the negative components. In prior studies \cite{keefe}, a partial set of positive exponents has been used to extract a polynomial fit, from which the entire spectrum was determined. Here, the negative components are directly evaluated in order to increase the accuracy of the results. The main disadvantage of this procedure is then that the range of Reynolds numbers that could be studied is vastly limited since the number of positive exponents is known to increase with the Reynolds number exponentially~\cite{constantin1985determining}. 

As a starting point, the structure of the exponent spectrum can be examined from the validation case investigated in Sec.~\ref{sec:numerics_lyap}. Figure~\ref{fig:LEvalid} shows the spectrum of the first 100 exponents. The structure of this plot is similar to that for other systems, such as turbulent channel flow \cite{keefe}. Overall, a finite set of positive exponents is observed, followed by a long tail of negative exponents. The magnitude is inversely related to the index, with near-linear scaling, as opposed to the K-S system, where negative exponents were found to scale as the fourth power of the Lyapunov index \cite{yang2009hyperbolicity}. This difference is likely due to the stronger dissipation term (fourth derivative) in the K-S system as opposed to the second-order viscous term in the Navier-Stokes equations. Further, the near-zero LEs show a knee-like structure, where the values do not follow the linear trend observed for the lower and higher LE indices. This structure has also been observed in Kolmogorov flow \cite{inubushi2012covariant} and Hamiltonian description of the motion of a collection of two-dimensional discs~\cite{bosetti2013orthogonal}.

\subsection{Scaling of the dimension of the attractor}
\label{sec:scaling}

The scaling of the dimension of the attractor of turbulent flows with the Reynolds number has important implications, in particular, to decide the resolution required to capture the dynamics of the flow field \cite{foias1988inertial,akram2019reduced,maryam_aim_jcp}. To estimate this scaling relation, a series of calculations with varying Reynolds numbers is conducted. Through the Kaplan-Yorke (KY) conjecture \cite{kaplan1979chaotic}, the geometric dimension of the system's attractor can be related to its LEs. The KY dimension is expressed as 
\begin{equation}
    D_{KY} = i + \frac{\sum_{1}^i \lambda_j}{ | \lambda_{i+1} |},
\end{equation}
where $\lambda_j$ denotes the $j^{th}$ LE, and $i$ is the last index such that $\sum_1^i \lambda_j \geq 0 $. The dimension of the attractor of the system can inform about the complexity of the dynamics considered as well as serve as an indicator that determines the minimal number of degrees of freedom required to capture all the dynamics on the attractor \cite{maryam_aim_jcp}. All calculations use the linear forcing techniques discussed in Sec.~\ref{sec:numerics_lyap}. The statistics of these cases are provided in Tab.~\ref{tab:statRe}.

\begin{table}
\caption{\label{tab:statRe} Turbulent statistics of the simulations conducted with various $Re_{\lambda}$ using the linear forcing scheme. $Re_{\lambda}$ denotes the Reynolds number based on the Taylor microscale; $\epsilon$ is the energy dissipation rate; $l_{int}$ is integral length scale based on the turbulent kinetic energy and the energy dissipation rate; $k$ is the turbulent kinetic energy; $N$ is the number of modes in one direction. $m$ is the number of LE computed; $A$ is the linear forcing coefficient; $D_{KY}$ is the computed Kaplan-Yorke dimension; $T/\tau$ is the total simulation time divided by the eddy turnover time.}
\begin{ruledtabular}
 \begin{tabular}{||c | c c c c||} 
 Case & 1 & 2 & 3 & 4\\ [0.5ex] 
 \hline
 $Re_{\lambda}$ & $15.55$ & $21.26$ & $25.57$ &  $37.67$ \\ 
 \hline
 $\epsilon$ [m$^2$/s$^3$]  & $0.07$  &  $0.26$  & $0.68$ & $8.8$ \\ 
 \hline
 $l_{int}$ [m] & $0.47$ & $0.53$ & $0.55$ & $0.52$ \\
 \hline
 $k$ [m$^2$/s$^2$] & $0.35$ & $0.93$ & $1.8$ & $9.7$\\
 \hline
 $N$ &  $32$ & $64$ & $64$ & $128$ \\
  \hline
 $m$ &  $99$ & $199$ & $299$ & $49$ \\
 \hline
 $A$ &  $0.137$ & $0.215$ & $0.3$ & $0.75$\\
 \hline
 $D_{KY}$ &  $57.67$ & $128.67$ & $232.54$ & $1430.7$\\
 \hline
 $T/\tau$ &  $743$ & $589$ & $546$ & $206$\\
 \hline
 $\nu$ &  $0.05$ & $0.05$ & $0.05$ & $0.05$\\
  \hline
 $\kappa_{max} \eta$ &  $1.1$ & $1.5$ & $1.2$ & $1.3$\\

\end{tabular}
\end{ruledtabular}
\end{table}

\begin{figure}[hb]
\begin{center}
\includegraphics[width=0.4\textwidth,trim={0cm 0cm 0cm 0cm},clip]{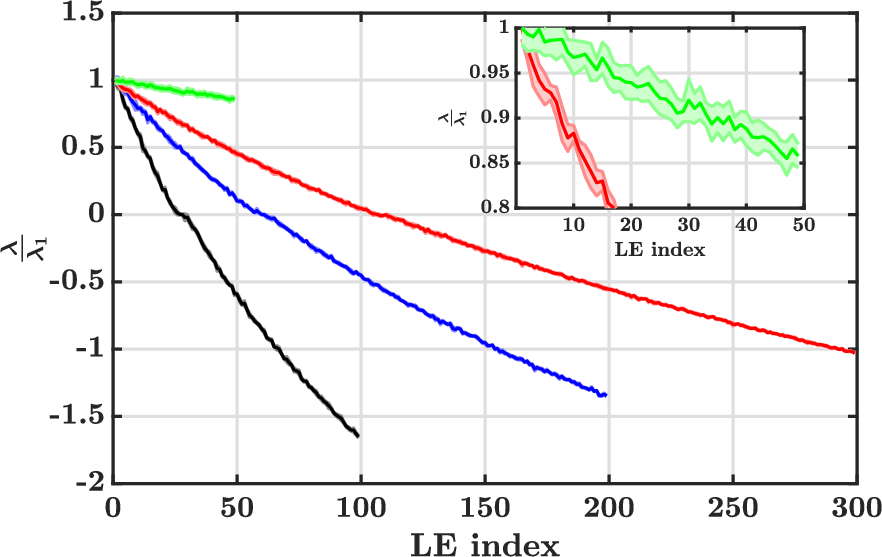}
\caption{LE rescaled by the first LE computed for Case 1 (\mythickline{black}), Case 2 (\mythickline{blue}), Case 3 (\mythickline{red}), Case 4 (\mythickline{green}). The shaded areas denote the statistical errors. Inset: Magnified region near the first exponents for Cases 3 and 4.}
\label{fig:LEhighRE}
\end{center}
\end{figure}

The LE computed for each one of these cases are shown in Fig.~\ref{fig:LEhighRE}. For cases 1-3, a sufficient number of exponents are obtained to estimate the dimension from the above relation directly. In practice, the exponents are obtained from long-time averages. Therefore, these values are subject to statistical errors, which may be estimated from the time-series of the exponents. The approach used here is based on the technique proposed in \cite{uncertaintyDNS_oliver} for turbulence statistics. For the average turbulence quantities ($Re_{\lambda}$, Kolmogorov length scale $\eta$), uncertainty estimates are obtained following the same procedure. Here, the simulations are run for several hundred eddy turnover times in order to minimize the statistical uncertainty. The simulation parameters are shown in Tab.~\ref{tab:statRe}. The resulting LE and corresponding statistical error are shown in Fig.~\ref{fig:LEhighRE}. An uncertainty estimate for the attractor dimension of Case 1, 2 and 3 can then be derived by computing $D_{KY}$ using the mean LE shifted by one standard deviation up or down.  For Case 4, the number of computed exponents is not sufficient to estimate the dimension of the attractor (the first 49 exponents are all positive). Since it is computationally expensive to obtain more exponents, $D_{KY}$ must be estimated using an extrapolation method for the LE. For this purpose, it is recognized that the shape of the Lyapunov spectrum is similar in all the cases considered and can be approximated as a power-law function of the Lyapunov index~\cite{hassanaly2019ensemble}. Since the extrapolation procedure can now affect the $D_{KY}$, it is necessary to estimate the uncertainty that it generates for the estimation of the attractor dimension. Note that the functional form of the spectrum can reasonably be assumed to be the same, but cannot be expected to use the same exponential decay rate as can be seen in Fig.~\ref{fig:LEhighRE}. The uncertainty quantification procedure is explained in detail in Appendix~\ref{sec:app_dimuq}.

\begin{figure}[h]
\begin{center}
\includegraphics[width=0.4\textwidth,trim={0cm 0cm 0cm 0cm},clip]{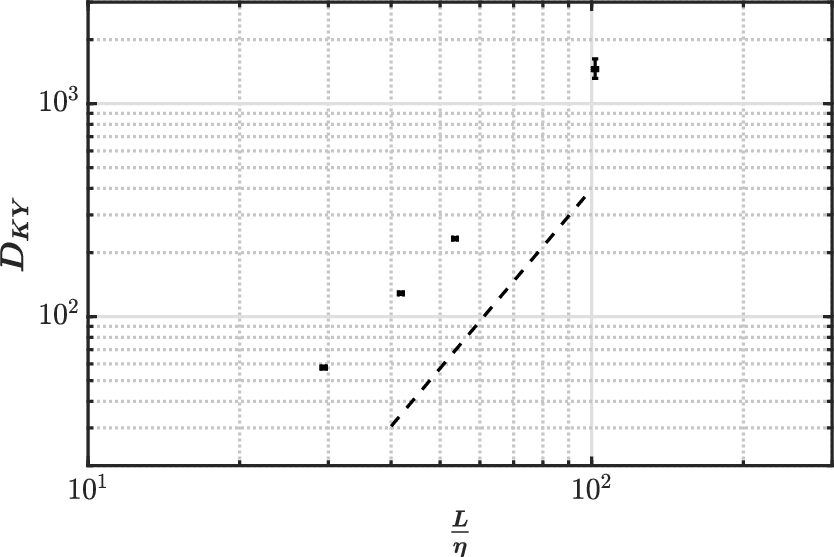}
\caption{Kaplan Yorke dimension obtained from Case 1, 2, 3 and 4 (variable $Re_{\lambda}$ and linear forcing scheme) plotted against the ratio $\frac{L}{\eta}$, along with a $(\frac{L}{\eta})^{2.8}$ slope (\mythickdashedline{black}). The error bars denote uncertainty estimate for both the x and y-axis due to statistical convergence and extrapolation uncertainty.}
\label{fig:scaling} 
\end{center}
\end{figure}

Prior estimates \cite{LLbook} show that the attractor dimension scales as $(\frac{L}{\eta})^3$, where $L$ denotes the length of the domain in one direction, and $\eta$ is the Kolmogorov length scale. Other refinements of the estimate have been mathematically derived \cite{constantin1985determining,gibbon1997attractor}, but involve the scaling with the upper bound rather than average turbulent flow quantities. Here, the scaling of the dimension with the length scale ratio is shown in Fig.~\ref{fig:scaling}. It is seen that $D_{KY} = (L/\eta)^{2.8 \pm 0.095}$, which is close to the theoretical estimate. The procedure to obtain the uncertainty estimate is described in Appendix~\ref{sec:app_dimuq}. The fact that the theoretical scaling is based on fully developed high Reynolds number turbulence, where the underlying assumptions regarding the separation of scales are valid, indicates that the attractor properties are only weakly dependent on these assumptions. 

\subsection{The response of the flow to perturbations}
\label{sec:lv}

\begin{figure}[hb]
\begin{center}
\includegraphics[width=0.4\textwidth,trim={0cm 3cm 0.1cm 5cm},clip]{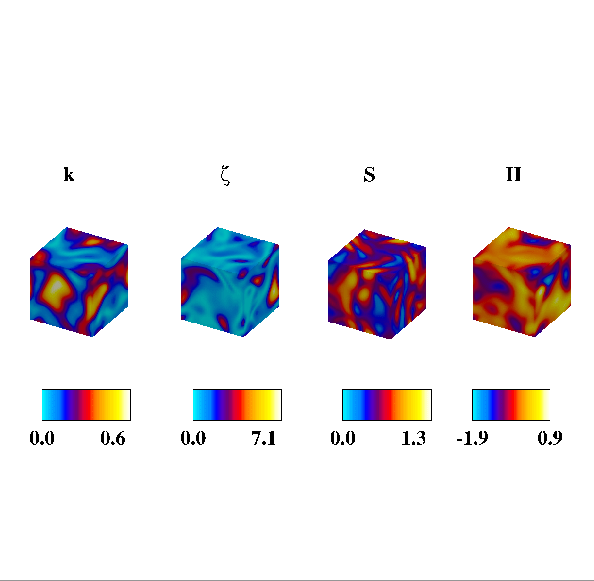}
\includegraphics[width=0.4\textwidth,trim={0cm 3cm 0.1cm 5cm},clip]{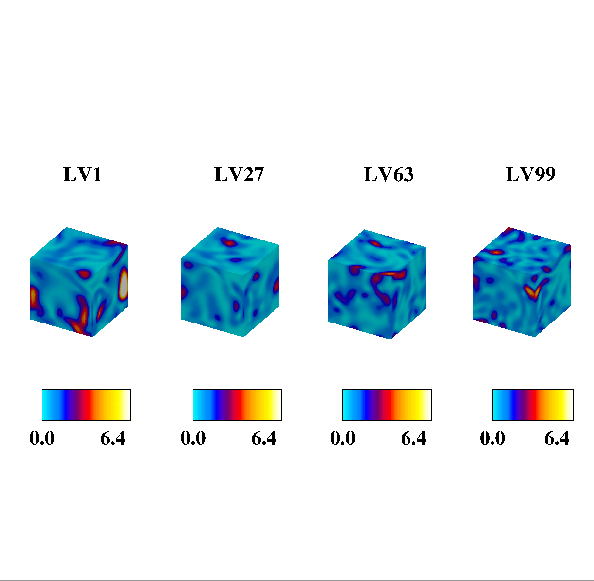}
\caption{Contours of Case 1 taken at the same instant. Top: instantaneous contour of turbulent kinetic energy $k$, enstrophy $\zeta$, strain rate $S$ and helicity density $H$ (from left to right). Bottom: contour of $\delta \bxi^2$ for the most chaotic first LV, the 27th LV corresponding to near-zero LE, the 63rd LV, and the 100th LV (from left to right).}
\label{fig:illustation} 
\end{center}
\end{figure}

So far, the discussion has centered on the Lyapunov exponents. In order to assess the correlation between the flow field and the Lyapunov perturbations, the Lyapunov vectors have to be studied. Each LV is a three-dimensional flow field made of three variables (one for each velocity component) at every grid point. To identify where perturbations grow the most, the energy of the LV $\delta \bxi^2$ can be computed as the sum of squares of the velocity components that compose the normalized LV and rescaled by a factor $1/N^3$, where $N^3$ is the total number of grid points. In Fig.~\ref{fig:illustation}, the first, 27th and the 100th LV obtained for Case 1 are plotted alongside the turbulent kinetic energy of the flow field $k$, its helicity density $H$, its enstrophy $\zeta$, and the strain rate magnitude $S$. Noting $\bx$ the physical space location, $\boldsymbol{u}$ the velocity field, the turbulent kinetic energy is defined as $k(\bx) = \frac{1}{2} \boldsymbol{u} \cdot \boldsymbol{u}$, enstrophy is defined as  $\zeta(\bx) = (\nabla \times \boldsymbol{u}) \cdot (\nabla \times \boldsymbol{u})$, the helicity density is defined as $H(\bx) = (\nabla \times \boldsymbol{u}) \cdot \boldsymbol{u}$, and the strain rate $S$ is defined as the Frobenius norm of the strain rate tensor $\frac{1}{2} (\frac{\partial u_i}{\partial x_j} + \frac{\partial u_j}{\partial x_i})$. From these plots, it is seen that the LVs themselves are disorganized fields, containing structures similar to the original velocity field. A detailed statistical analysis of the correlation between the LVs and the flow field quantities considered is performed and discussed below. To understand the correlation between the flow and the LVs, statistical analysis is performed and is discussed below.

\subsubsection{Localization of chaotic response}
\label{sec:local}

It has been reported \cite{forster2004perturbed,xu2016covariant} in different chaotic systems that only a small part of the physical space was responsible for most of the perturbation growth in chaotic systems. This property is called the localization. In the case of hard-disc systems, it has been found that the number of particles that contribute to perturbation growth decreased as the total number of particles in the system increases \cite{forster2004perturbed}. This finding led to speculations about the convergence of the LS in the thermodynamic limit (with a large number of particles). In fluid systems, it has also been found that the most chaotic LV could be highly localized in physical space and that the least chaotic LV could be more spatially distributed \cite{egolf2000mechanisms}. Here, chaotic LV refers to the vector associated with a positive LE. The most chaotic LV is the first LV, and the least chaotic LV is the last LV computed. As a side note, the localization has been mostly investigated with GSV but has also been recently examined using CLV \cite{morriss2012localization}, where it was observed that while the spatial distribution of the least chaotic LV was less pronounced for the CLV than the GSV, these were still more distributed than the most chaotic CLV. In this section, the localization of the LV is the main focus. Here, the variation of the localization with respect to the turbulence level, the Lyapunov index and other macroscopic properties such as local kinetic energy are investigated. 

\begin{figure}[hb]
\begin{center}
\includegraphics[width=0.4\textwidth,trim={0cm 0cm 0cm 0cm},clip]{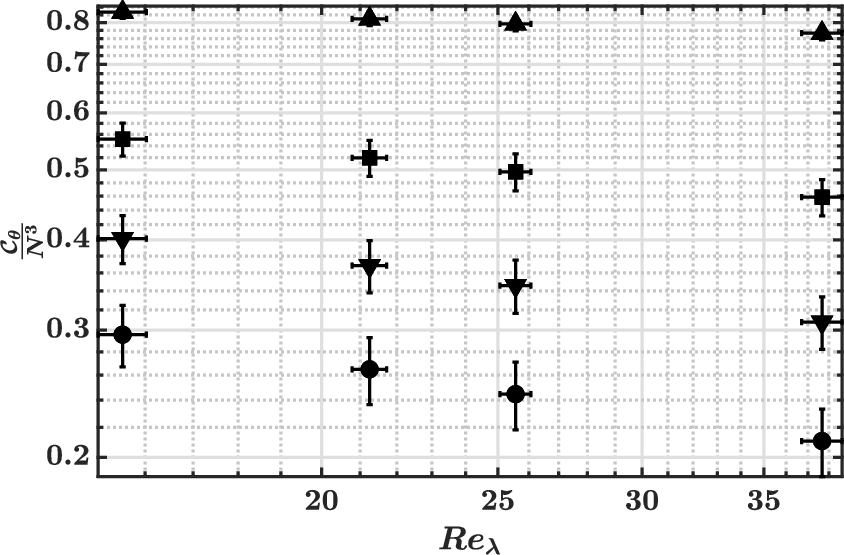}
\includegraphics[width=0.4\textwidth,trim={0cm 0cm 0cm 0cm},clip]{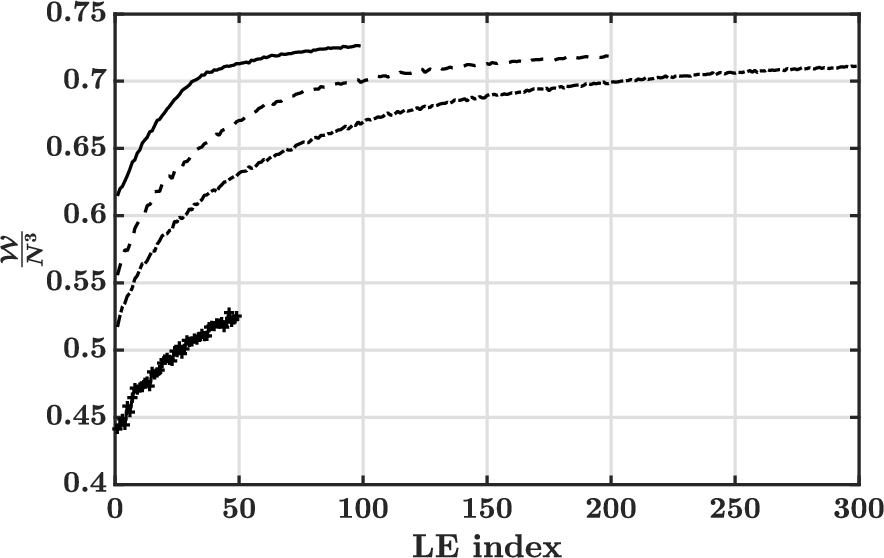}
\caption{Physical space localization of the LV. Top: time average localization $C_{\theta}$ obtained from Case 1, 2, 3 and 4 with different threshold $\theta = 0.98$ (\mythicktriangle{black}{black}), $\theta = 0.88$ (\mythicksquare{black}{black}), $\theta = 0.78$ (\mythickVtriangle{black}{black}), $\theta = 0.68$ (\mythickcircle{black}{black}). The vertical error bars denote the RMS fluctuations around the mean (as opposed to statistical uncertainty). The horizontal error bars denote statistical uncertainty in the average value of $Re_{\lambda}$. Bottom: localization width plotted against the LE index obtained from Case 1 (\mythickline{black}), 2 (\mythickdashedline{black}), 3 (\mythickdasheddottedline{black}) and 4 (\mybarredcross{black}).}
\label{fig:localization} 
\end{center}
\end{figure}

The first mathematical definition of the LV localization is inspired by Ref.~\cite{forster2004perturbed}. Here, the parameter $C_{\theta}$ is defined by counting the number of vectors entries of $\delta \bxi^2$ that contribute to a certain fraction $\theta$ of its total L$_2-$norm. More formally, let $\delta \boldsymbol{\Gamma}^2$ be defined as $\delta \bxi^2$ with entries sorted in descending order. Then $C_{\theta}=j$, where $j$ is such that $\sum_{i=1}^{j} \delta \bg^2_i \geq \theta \norm{\delta \bxi}^2$ and $\sum_{i=1}^{j-1} \delta \bg^2_i < \theta \norm{\delta \bxi}^2$. In Fig.~\ref{fig:localization}~(top), the localization of the most chaotic LV is shown as a function of $Re_{\lambda}$ for different thresholds $\theta$. It appears that the fraction of the domain that contributes to the perturbation growth gets smaller and smaller as the level of turbulence increases. This result suggests that tracking the evolution of perturbations over time using measurement techniques to anticipate their chaotic build-up would get more and more difficult as the Reynolds number increases. In turn, if one wants to exploit the chaoticity of the flow field for flow control purposes, the localized response of the flow field could facilitate targeted control for high-Re flows. At the moment, while it is unclear whether this trend would continue for higher levels of turbulence, it can be reasonably expected to follow the rate at which Kolmogorov scales decreases with increasing $Re_{\lambda}$.

The variation of the localization as a function of the LE index is then examined. To do so, the entropy-like metric of localization introduced in Ref.~\cite{taniguchi2003localized} is used. This metric $\frac{\mathcal{W}}{N^3}$, called the \textit{localization width}, where $N^3$ is the number of entries in $\delta \bxi^2$ and is defined as $\mathcal{W}=exp(S)$, where $S = - \langle \sum_{j=1}^{N^3} \delta \xi^2_j~\text{log}~\delta \xi^2_j \rangle$, and $\delta \xi^2_j$ is the $j$-th entry of $\delta \bxi^2$ normalized by the L$_2$-norm of $\delta \bxi^2$. This metric is advantageous since it can be easily computed as it does not require sorting the entries of $\delta \bxi^2$, and it does not depend on the value of a particular threshold. However, this metric is bounded between $\frac{1}{N^3}$ and $1$, and its value depends on the total number of vector entries (number of grid points): the same value of $\frac{\mathcal{W}}{N^3}$ for cases discretized with different number of modes can mean that a field is localized in one case and distributed in the other. As opposed to $C_{\theta}$, this metric is not suited for comparing fields discretized with a different number of grid points which explains why it was not used to compare different $Re_{\lambda}$ simulated with different numbers of Fourier modes. Fig.~\ref{fig:localization}~(right) shows the localization width obtained for Cases 1, 2, 3 and 4. Here, the statistical uncertainty of the quantity is not indicated for clarity. Similar to previous results mentioned at the beginning of this section, the LVs are increasingly distributed in physical space as the LE index increases, although the range of variation is narrower than in other systems \cite{taniguchi2003localized}.

The metrics used above characterize the level of localization of the perturbations but do not indicate the location where perturbations grow the most. To answer this question, the conditional averages of $\delta \bxi^2$ with turbulent flow quantities are examined. Below, only the results from the linear forcing technique based simulations are used, since the results were found consistent across different forcing schemes (See Supplementary material).

\begin{figure}[hb]
\begin{center}
\includegraphics[width=0.4\textwidth,trim={0cm 0cm 0cm 0cm},clip]{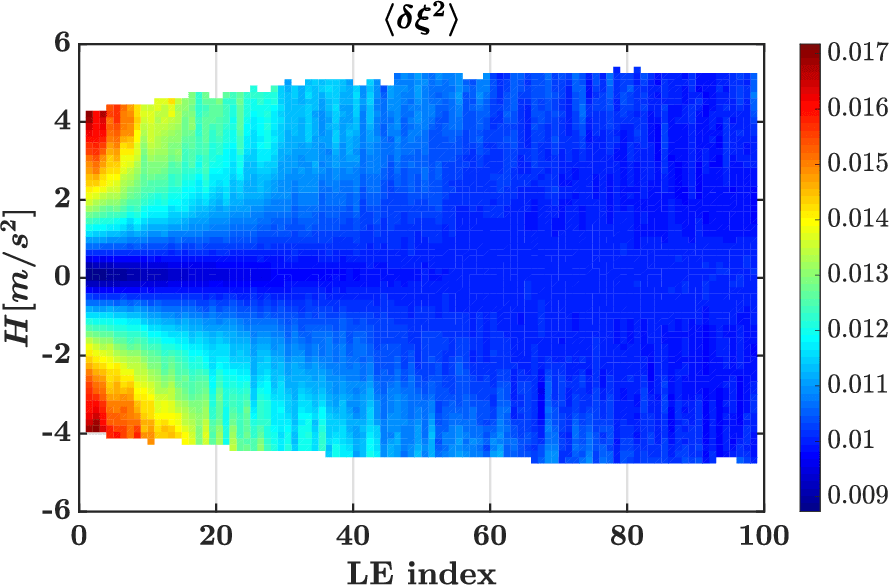}
\includegraphics[width=0.4\textwidth,trim={0cm 0cm 0cm 0cm},clip]{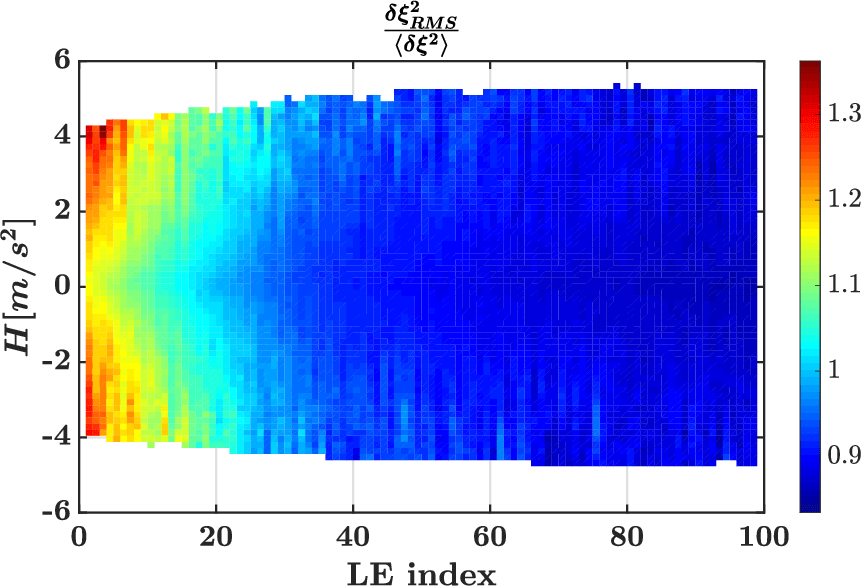}
\caption{Top: conditional average of $\delta \bxi^2$ conditioned on helicity density values $H$, for each LV. Bottom: conditional RMS of $\delta \bxi^2$ conditioned on helicity density values $H$, for each LV, rescaled by the conditional average of $\delta \bxi^2$.}
\label{fig:correlationHE} 
\end{center}
\end{figure}

First, the conditional average of $\delta \bxi^2$ conditioned on the helicity density $H$ is examined (Fig.~\ref{fig:correlationHE}) for Case 1. Note that only the points with relative statistical uncertainty lower than $1\%$ are plotted here. It is seen that large values of $\delta \bxi^2$ are correlated with large absolute values of helicity density for the most chaotic LV only. This suggests that perturbations grow where helicity density is the largest. In turn, large values of the dissipative LV appear uncorrelated with helicity density. Figure~\ref{fig:correlationHE} also shows the conditional root mean square (RMS) of $\delta \bxi^2$. This data shows that at high helicity density values, the variation in the LV energy is also high. This suggests that although high helicity density regions are associated with increased perturbation growth, this feature is not persistent and there are times when the growth is small. In fact, the range of variation exceeds the conditional average.

\begin{figure}[hb]
\begin{center}
\includegraphics[width=0.4\textwidth,trim={0cm 0cm 0cm 0cm},clip]{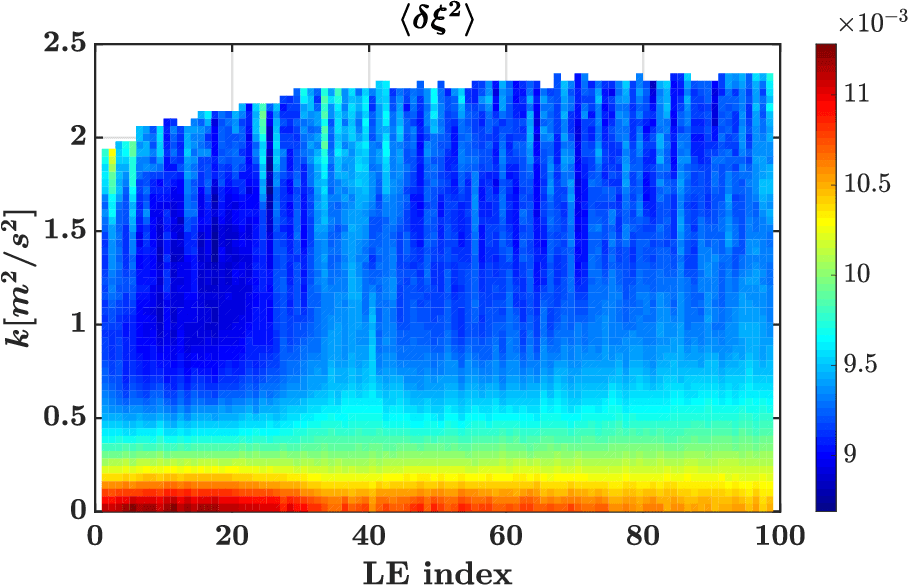}
\includegraphics[width=0.4\textwidth,trim={0cm 0cm 0cm 0cm},clip]{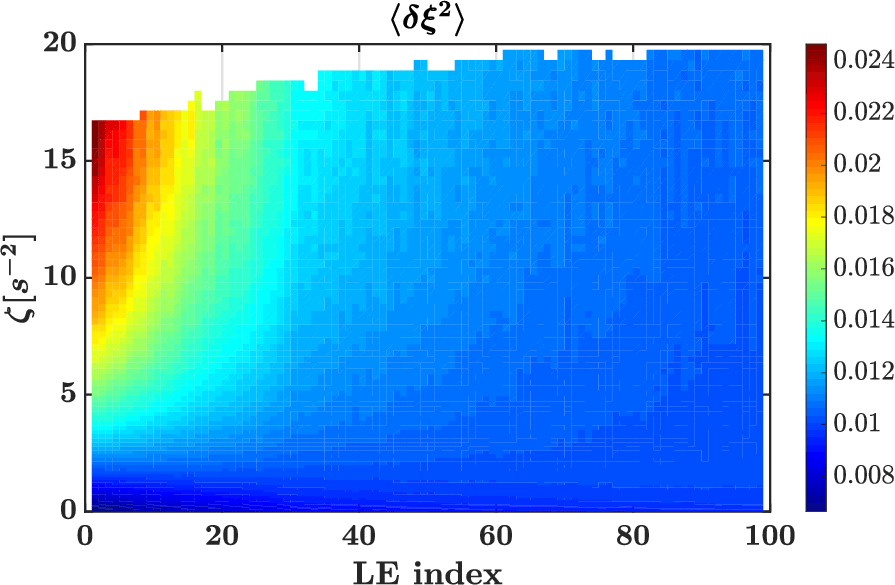}
\includegraphics[width=0.4\textwidth,trim={0cm 0cm 0cm 0cm},clip]{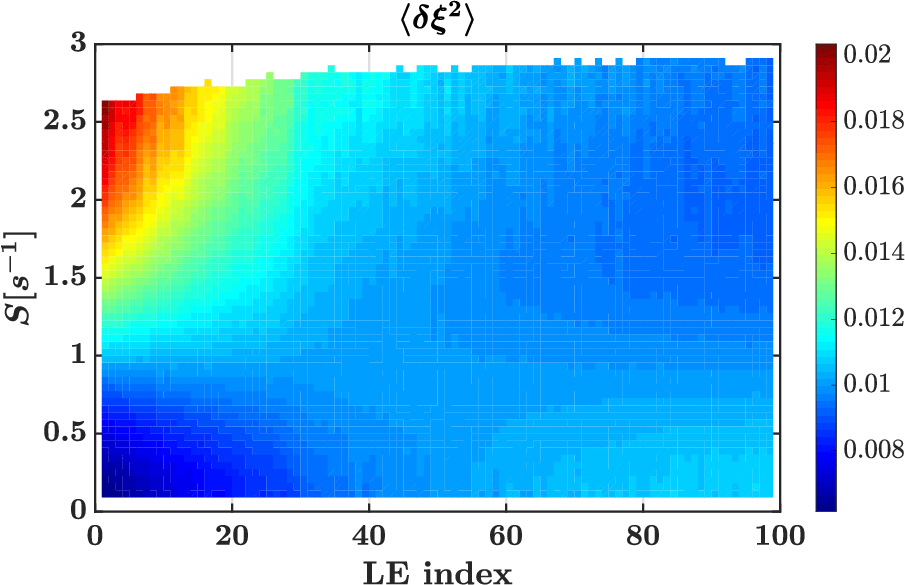}
\caption{Top: conditional average of $\delta \bxi^2$ conditioned on turbulent kinetic energy values $k$, for each LV. Middle: conditional average of $\delta \bxi^2$ conditioned on enstrophy values $\zeta$, for each LV. Bottom: conditional average of $\delta \bxi^2$ conditioned on strain rate values $S$, for each LV. Results correspond to Case 1.}
\label{fig:correlationKEVort}
\end{center}
\end{figure}

The local helicity density can be intuitively understood as containing a contribution stemming from the local kinetic energy and from the local enstrophy. It is then natural to investigate the conditional averages with both fields to identify the importance of each component on the localization of the LV. Figure~\ref{fig:correlationKEVort} (left) shows the conditional averages of $\delta \bxi^2$ with $k$. As opposed to the findings obtained with helicity density, the first GSV (which is the first CLV) suggests that perturbations grow where the local kinetic energy is low. In turn, the conditional averages with enstrophy (right) are similar to the ones observed with helicity density with slightly higher conditional averages. The conditional average with respect to strain rate is shown in Fig.~\ref{fig:correlationKEVort}. It can be seen that conditioning on strain rate produces higher averages as compared to conditioning on helicity but lower values compared to enstrophy.  While enstrophy can be considered to be a better marker for isolating the most chaotic locations in the domain than strain rate and helicity, the fact that the conditional averages with enstrophy, strain rate and helicity are close suggests that perturbations expand mostly in regions of large velocity gradients.

\begin{figure}[hb]
\begin{center}
\includegraphics[width=0.4\textwidth,trim={0cm 0cm 0cm 0cm},clip]{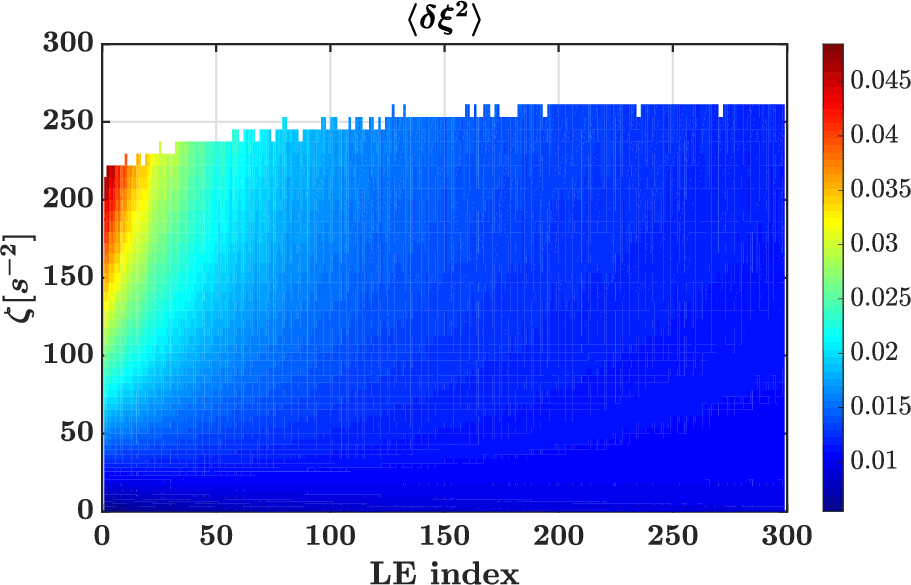}
\includegraphics[width=0.4\textwidth,trim={0cm 0cm 0cm 0cm},clip]{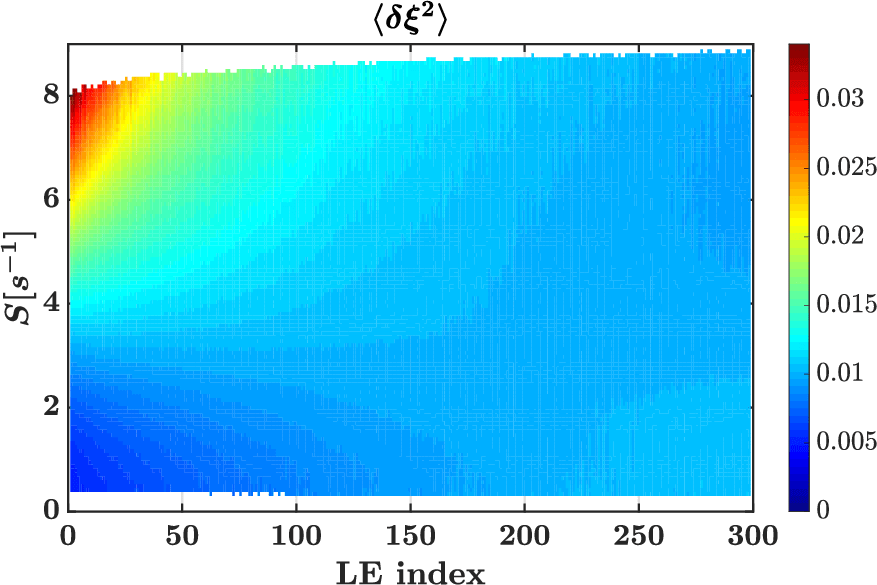}
\caption{Top: conditional average of $\delta \bxi^2$ conditioned on enstrophy values $\zeta$, for each LV. Bottom: conditional average of $\delta \bxi^2$ conditioned on strain rate values $S$, for each LV. Results correspond to Case 3.}
\label{fig:higheREconditionalAv}
\end{center}
\end{figure}

Figure~\ref{fig:higheREconditionalAv} shows that, at higher Reynolds numbers, $\langle \delta \bxi^2 | \zeta \rangle $ exceeds $\langle \delta \bxi^2 | H \rangle $ and $\langle \delta \bxi^2 | S \rangle $ by a larger amount compared to the same averages at lower Reynolds number. In other words, enstrophy becomes a better marker with increase in Reynolds number.  Only the results of Case 3 are shown for the sake of brevity. The data from Cases 2 and 4 and the comparison with averages conditioned on helicity are available as supplementary material.

\begin{figure}[hb]
\begin{center}
\includegraphics[width=0.4\textwidth,trim={0cm 0cm 0cm 0cm},clip]{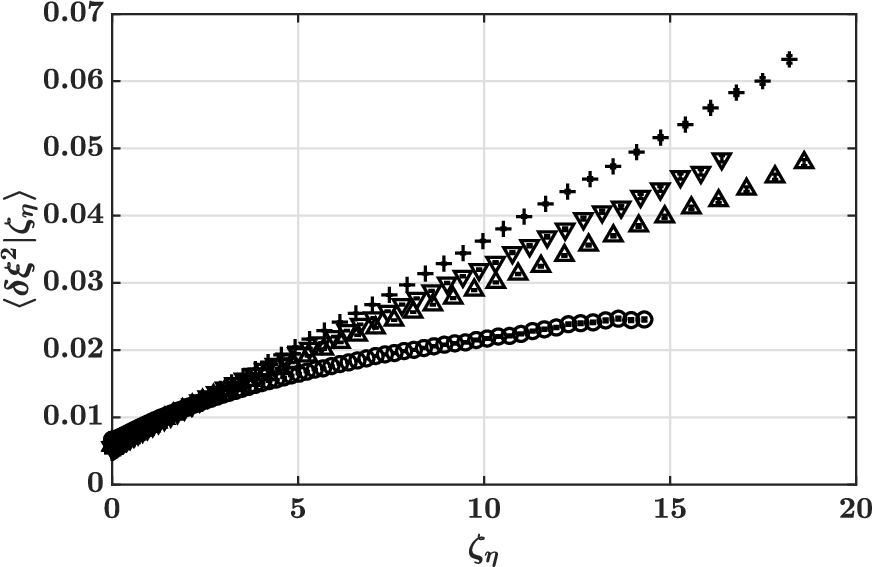}
\caption{Conditional average of $\delta \bxi^2$ at different rescaled enstrophy $\zeta_n$ for Case 1 (\mythickcircle{black}{white}), Case 2 (\mythicktriangle{black}{white}), Case 3 (\mythickVtriangle{black}{white}), Case 4 (\mythickcross{black}). Error bars show the statistical uncertainty.}
\label{fig:enstrophyhighRE}
\end{center}
\end{figure}

The average values of $\delta \bxi^2$ for the first LE index, conditioned on enstrophy can also be compared across different Reynolds numbers to investigate its effect on the findings listed above. Here, a normalized enstrophy $\zeta_{\eta} = \zeta \tau_{\eta}^2$ is constructed, where $\tau_{\eta}$ is the Kolmogorov time scale. Note that microscale scaling is adopted given than $\zeta$ is a gradient-based quantity. Figure~\ref{fig:enstrophyhighRE} shows the obtained conditional average results, along with the statistical uncertainties. The trends noted in Fig.~\ref{fig:correlationKEVort} hold across the Reynolds number considered. Overall, the large values of $\zeta_{\eta}$ are a better marker for the chaotic response of the flow to perturbations as the Reynolds number increases. 

\begin{figure}[hb]
\begin{center}
\includegraphics[width=0.4\textwidth,trim={0cm 0cm 0cm 0cm},clip]{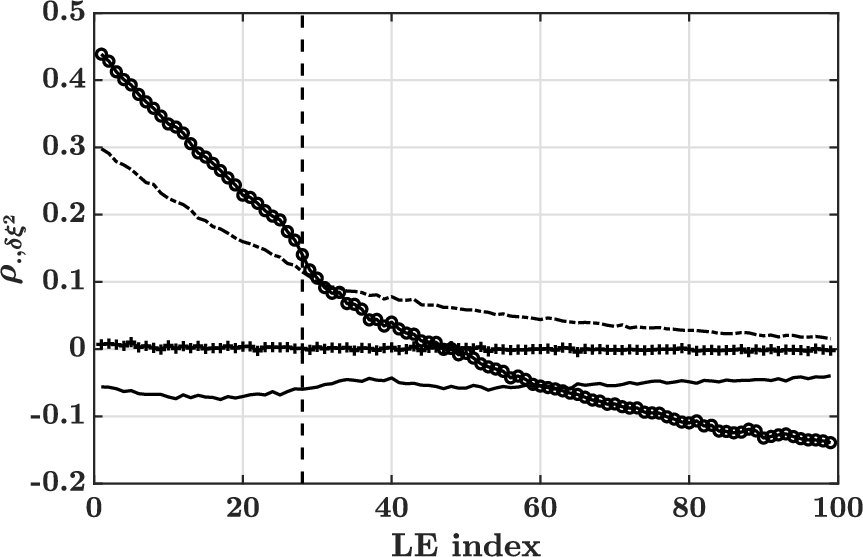}
\caption{Field correlation of $\delta \bxi^2$ and $\zeta$ (\mythickdasheddottedline{black}), $\delta \bxi^2$ and $H$ (\mybarredcross{black}), $\delta \bxi^2$ and $k$ (\mythickline{black}), $\delta \bxi^2$ and $S$ (\mybarredcircle{black}{white}). The curves are plotted alongside the index where the LS crosses zero (\mythickdashedline{black}).}
\label{fig:correl}
\end{center}
\end{figure}

To further characterize the spatial structure of the perturbation growth, statistical correlations between different variables are used. The correlation of two fields $\phi$ and $\psi$ is defined as 

\begin{equation}
    \rho_{\phi,\psi}= \frac{(\phi - \langle \phi \rangle ) \cdot (\psi - \langle \psi \rangle ) }{ \norm{\phi - \langle \phi \rangle }   \norm{\psi - \langle \psi \rangle } }, 
\end{equation}
where the norm considered is the L$_2$-norm and $\langle \cdot \rangle$ denotes a spatial average. Figure~\ref{fig:correl} shows the field correlation of $\delta \bxi^2$ with enstrophy, helicity density, turbulent kinetic energy and strain rate. In line with the findings of Sec.~\ref{sec:local}, the LV appears slightly anticorrelated with the turbulent kinetic energy, where low values of kinetic energy provide high localized energy. Since helicity density is symmetric about the zero value, the correlation with $\delta \bxi^2$ is close to zero. Only enstrophy and strain rate show significant correlation with the chaotic LVs. For Case 1, strain rate appears to be the turbulent quantity that describes best the structures of $\delta \bxi^2$. However, at higher Reynolds numbers, the chaotic LVs become more correlated with enstrophy. This result is illustrated in Appendix~\ref{app:correlationHigherRe} and echos the previous observation obtained with conditional averages.

\begin{figure}[hb]
\begin{center}
\includegraphics[width=0.4\textwidth,trim={0cm 0cm 0cm 0cm},clip]{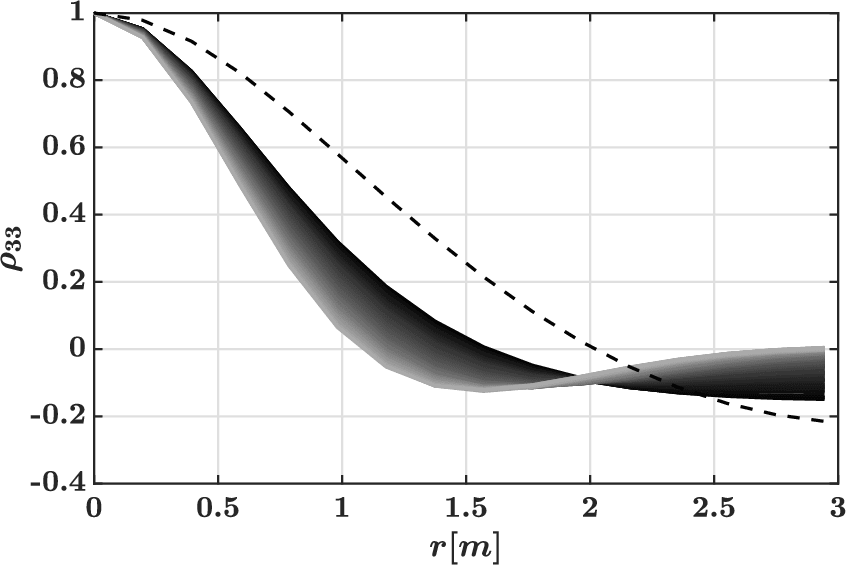}
\includegraphics[width=0.4\textwidth,trim={0cm 0cm 0cm 0cm},clip]{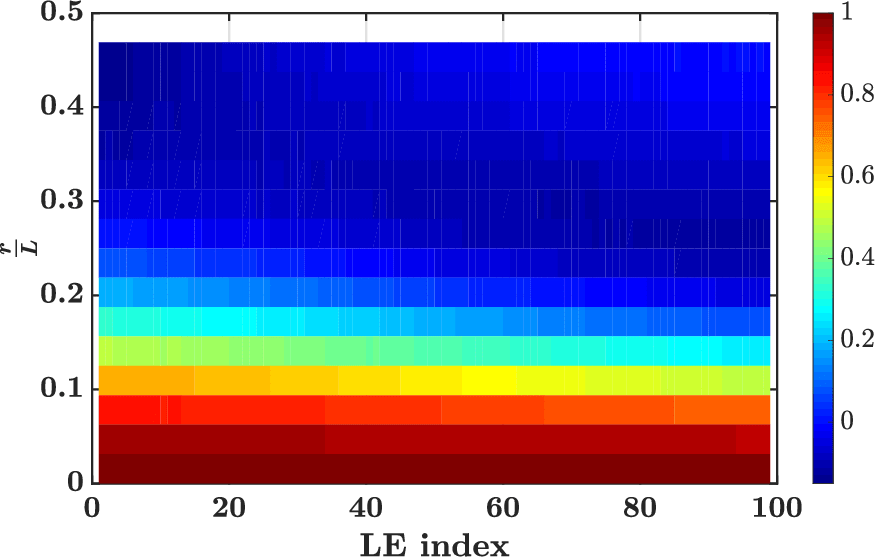}
\caption{Top: spatial correlation $\rho_{33}(r)$ of the LV (solid lines) and of the underlying flow-field (\mythickdashedline{black}). The lighter the solid line, the higher the Lyapunov index.} Bottom: spatial correlation  $\rho_{33}(r)$ of the LV plotted against the LE index and the distance. Both plots are generated with the data of Case 1. $L$ denotes the box length ($2\pi$).
\label{fig:spatialcorrel}
\end{center}
\end{figure}

The two-point correlation for the flow field and the LVs are shown in Fig.~\ref{fig:spatialcorrel}. Here, the spatial correlation of $\bxi$ in the $x$ direction is given by 
\begin{equation}
    \rho_{ij}(r)= \frac{ \langle \bxi_i(\bx) \bxi_j (\bx+(r,0,0)) \rangle }{\norm{\bxi}^2}, 
\end{equation}
and is defined similarly for the underlying flow field. As expected, the integral length scale of the LV is smaller than that of the underlying flow field. However, there is slight reduction of integral length scale as a function of LE index, seen from the more rapid decorrelation for higher LE indices.

\begin{figure}[hb]
\begin{center}
\includegraphics[width=0.4\textwidth,trim={0cm 0cm 0cm 0cm},clip]{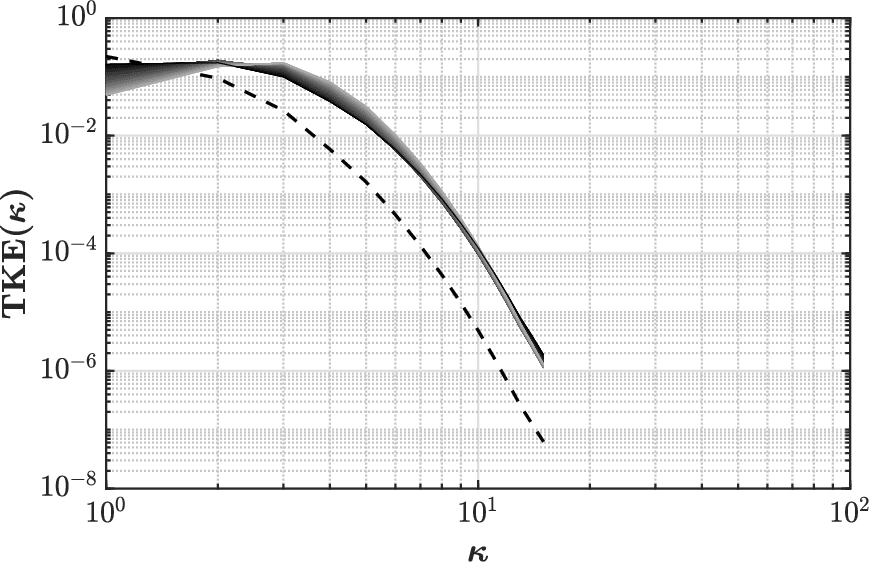}
\caption{Time-averaged energy spectrum of the computed LV for Case 1 (solid lines). Time-averaged energy spectrum of the underlying flow field (\mythickdashedline{black}). The lighter the solid line, the higher the Lyapunov index.}
\label{fig:spectrum} 
\end{center}
\end{figure}

The two-point correlation may be expressed in the spectral domain as the energy spectrum, which is shown in Fig.~\ref{fig:spectrum}. It is seen that there is a significant difference in the structure of the LV and the flow field spectra, with more energy at small scales observed for the Lyapunov fields. Further, the peak of the spectrum is located at larger wavenumbers, which is consistent with the two-point correlation (Fig~\ref{fig:spatialcorrel}). It is noted that this result differs from previous findings for the Kuramoto-Sivashinsky Equation (KSE) \cite{hassanaly2019numerical,yang2009hyperbolicity} in several aspects. For the KSE, the spectra of the chaotic and dissipative LV were found to differ significantly. Further, the spectrum of dissipative LV was found to be localized in Fourier space. Other work with the Rayleigh-B\'{e}nard convection showed that the energy spectra of the CLV were not independent of the Lyapunov indices \cite{xu2016covariant}. Given the present result, it can be expected that the first few CLV also have a non-localized energy spectrum. This analysis is presented in Appendix~\ref{sec:app_clv}.

\section{Summary and conclusions}
\label{sec:conclusion}

The present work computed for the first time, for a three-dimensional homogeneous isotropic flow, all the positive LE and LV of forced HIT flows for a set of $Re_{\lambda}$, while inferring the negative values for the highest Reynolds number case. The analysis of the LV revealed that the fraction of the spatial volume where the perturbations grow decrease with an increase in $Re_{\lambda}$. It was found that there exists a strong correlation between the chaotic parts of the domain and the velocity gradients at low Reynolds number. At higher Reynolds numbers, enstrophy becomes a good marker of chaoticity. It was found that the dimension of the attractor scales as $(\frac{L}{\eta})^{2.8}$ where $L$ is the domain size and $\eta$ is the Kolmogorov length scale, which is close to other estimates predicted by prior theoretical work. However, the actual dimension is much smaller than the full dimension of the phase-space, indicating that reduced-order models that capture the dynamics of the flow field, could be developed. 

The results also indicate that the evolution of perturbations in a Navier-Stokes based flow is considerably different than in other canonical systems, such as those studied using the KS equations. In particular, the Lyapunov vectors did not show significant localization in Fourier space with an increase in the index. This difference is likely due to the very strong diffusion term in the KS equation (fourth-order derivative) as opposed to the second-order viscous dissipation in fluid flow. This is also supported by the fact that the different forcing schemes used to sustain the turbulent were found to only mildly affect the results obtained about the LV (See Supplementary material). 

This work provided a comprehensive description of the properties of the Lyapunov spectrum that can be expected in three dimensional turbulence. These properties could be used to design or select models able to better capture the flow dynamics rather than spatial or temporal statistics. Ultimately, the Lyapunov exponents and vectors could be used to guide better the design of models that would be more responsive to perturbations and capture extreme and rare events. These extensions are being considered currently \cite{hassanaly2019computational}.




\appendix

\section{Size of perturbations}
 \label{app:perturbationSize}
 
The Lyapunov analysis is valid only for small perturbations to the flow. Past this regime, a saturation of perturbations can be observed \cite{ding2007nonlinear}. It is therefore necessary to ensure that the size of the perturbations remains small, even after they have exponentially grown. Table~\ref{tab:pertMag} provides the average magnitude of velocity perturbation per cell in the domain when the perturbations are initialized and after they have grown at the rate of the largest LE. The perturbations are quantified in terms of average kinetic energy per cell (respectively $k_{pert}^{init}$ and $k_{pert}^{final}$) and should be compared to the kinetic energy of the baseline flow. These values are obtained by assuming that the perturbation is uniformly distributed across the cells in the domain. As can be seen, the perturbations are small compared to the baseline flow.

\begin{table}
\centering
\caption{\label{tab:pertMag} Perturbation magnitude per cell for Case 1,2,3,4. The perturbation is assumed to be evenly distributed across 10\% of the domain; $u_{turb}$ is the turbulent velocity.}
 \begin{tabular}{||c | c c c c||}
  \hline
  Case & 1 & 2 & 3 & 4\\ [0.5ex] 
 \hline
  $k_{pert}^{init}$ [m$^2$/s$^2$] & $1.87e^{-12}$ & $2.33e^{-13}$ & $2.33e^{-13}$ &  $2.92e^{-14}$ \\ 
 \hline
  $k_{pert}^{final}$ [m$^2$/s$^2$]  & $4.99e^{-12}$ & $8.97e^{-13}$ & $1.23e^{-12}$ &  $5.66e^{-13}$ \\ 
 \hline
 $k$ [m$^2$/s$^2$] & $0.35$ & $0.93$ & $1.8$ & $9.7$ \\
 \hline
\end{tabular}
\end{table}

\section{Spatial correlations at higher Reynolds numbers}
\label{app:correlationHigherRe}

\begin{figure}[hb]
\begin{center}
\includegraphics[width=0.22\textwidth,trim={0cm 0cm 0cm 0cm},clip]{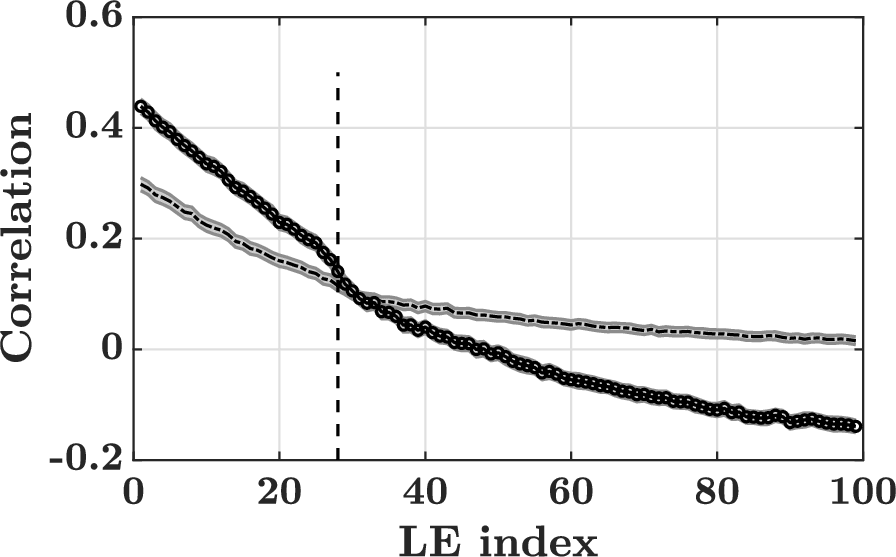}
\includegraphics[width=0.22\textwidth,trim={0cm 0cm 0cm 0cm},clip]{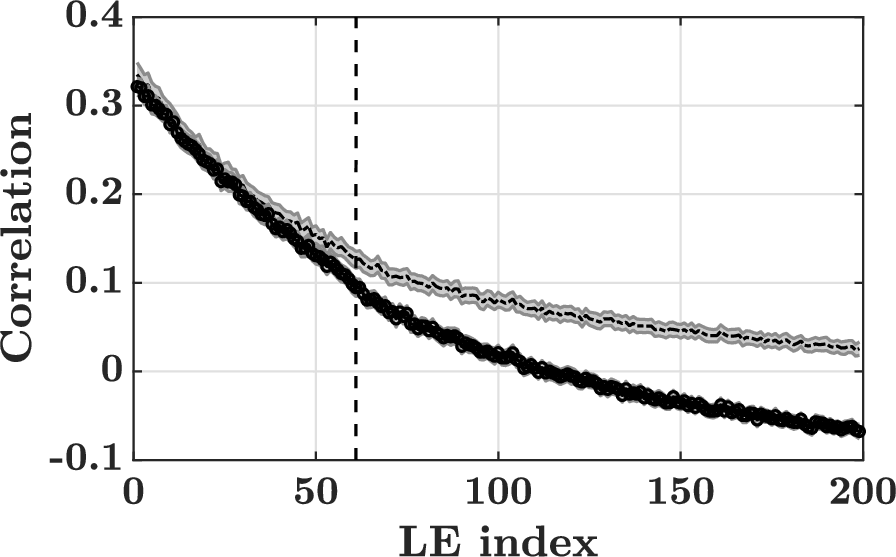}
\includegraphics[width=0.22\textwidth,trim={0cm 0cm 0cm 0cm},clip]{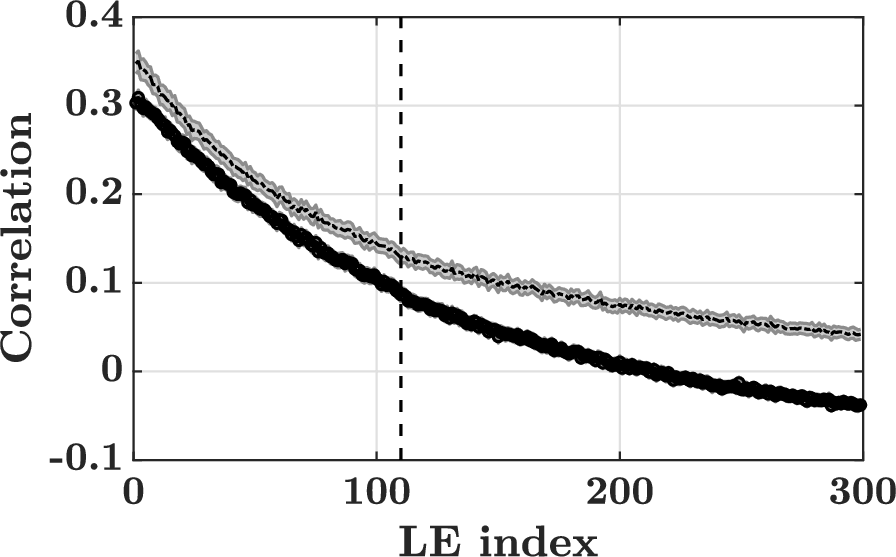}
\includegraphics[width=0.22\textwidth,trim={0cm 0cm 0cm 0cm},clip]{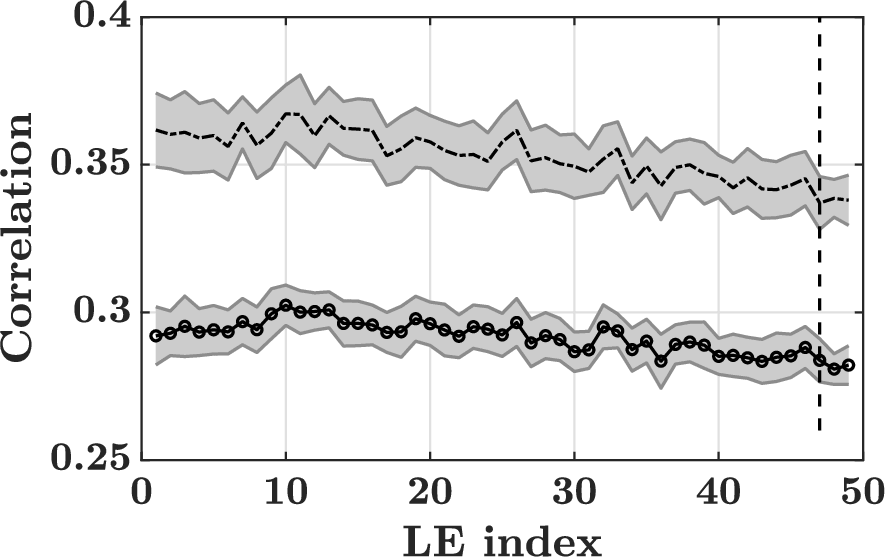}
\caption{Field correlation of $\delta \bxi^2$ and $\zeta$ (\mythickdasheddottedline{black}), $\delta \bxi^2$ and $S$ (\mybarredcircle{black}{white}). The curves are plotted alongside the index where the LS crosses zero (\mythickdashedline{black}). The shaded areas indicate the statistical errors. Top left: Case 1. Top right: Case 2. Bottom left: Case 3. Bottom right: Case 4.}
\label{fig:correlationStrainVort}
\end{center}
\end{figure}

Here, the effect of Reynolds number on correlations between the LVs and different flowfield quantities are presented. In Fig.~\ref{fig:correlationStrainVort}, it can be seen that the strain rate becomes less correlated with the most chaotic LVs as the Reynolds number increases. On the other hand, at high Reynolds numbers, enstrophy describes better  the spatial structure of the $\delta \bxi^2$. The computed statistical uncertainty suggests that this result is statistically significant.

\section{Effect of forcing scheme}
 \label{sec:app_forcingLE}
 
 In the case of forced HIT, despite the ability to run detailed simulations (resolve the smallest length scales), the turbulence needs to be sustained using an external volumetric force that compensates the dissipation. The functional form of this forcing term is a modeling choice that can have an impact on the turbulent flow field. Multiple forcing schemes have been proposed over the past decades with various objectives \cite{siggia1978intermittency,eswaran,carroll2013proposed,ghosal-dlm,palmore2018technique}; more recent techniques have focused on minimizing the time to statistical stationarity \cite{carroll2013proposed}. As a result, the forcing functions might, on purpose, alter the dynamics of the flow. Since the goal here is to study such dynamical aspects, it is necessary to understand the role of the forcing method on the results. 

Four different forcing schemes are considered. All of the cases are run using the spectral method described in Sec.~\ref{sec:numerics} and with $32^3$ modes. The following forcing methods are used :
\begin{itemize}
    \item Case 1 uses a linear forcing technique for all the wavenumbers, similar to \cite{Rosales:2005fy} and \cite{lundgren_linearforcing} with a linear forcing coefficient of $A=0.137$.
    \item Case 5 uses a linear forcing technique for only the lowest wavenumbers, similar to \cite{sullivan1994deterministic}. The linear forcing coefficient applied is $A=0.14$ for all wavenumbers with $|\kappa| \leq 3$.
    \item Case  6 uses a stochastic forcing technique \cite{eswaran}. The forcing functional form is based on an Ornstein-Uhlenbeck process. The parameters used are $T_L=0.92$, $K_f=2\sqrt{2}$ and $\varepsilon^* = 0.0015$. More precisely, the forcing term has the form 
    \[ \boldsymbol{f}(\kappa,t+\Delta t)=  (1-\frac{\Delta t}{T_L})\boldsymbol{f}(\kappa,t) + \boldsymbol{n}(\kappa) \sqrt{2 \varepsilon^* \frac{\Delta t}{T_L^2}}, \]
    for $|\kappa|\leq K_f $, where $\boldsymbol{n}(\kappa)$ is a complex number drawn from a standard normal distribution.
    
    Since the forcing term is not solely dependent on the position of the system in phase-space, but also depends on the random number drawn, the Lyapunov calculation is conducted by communicating, at every timestep, the same forcing term to all the forward realizations. In other terms, the same value for the forcing is used for the computation of each LE. The LEs are computed by using the ``same noise realization" technique in the sense of ``noise on the particle" (See \cite{laffargue2016lyapunov} for a detailed discussion).
    \item Case 7 is not a realistic forcing method but rather a numerical experiment used to compare Case 1, 5 and 6 in a fair manner. It consists of using a classical linear forcing technique with linear forcing coefficient $A=0.137$ for the unperturbed simulation (similar to Case 1) and communicating the same forcing to all the other perturbed simulations. This procedure is similar to the one used for Case 6.
    
\end{itemize}

The Reynolds number and statistics of all the cases are provided in Tab.~\ref{tab:statForc}, where $Re_{\lambda}$ is the averaged Reynolds number based on the Taylor microscales, $\epsilon$ is averaged the turbulent dissipation rate, $k$ is the averaged turbulent kinetic energy ($k$) and $l_{int}$ is the integral length scale. As can be seen, the statistics of the cases are similar for all forcing techniques except Case 6, which shows higher kinetic energy.

\begin{table}
\caption{\label{tab:statForc} Turbulent statistics of the simulations conducted with various forcing schemes.}
\begin{ruledtabular}
 \begin{tabular}{||c | c c c c||} 
 Case & 1 & 5 & 6 & 7 \\ [0.5ex] 
 \hline
 $Re_{\lambda}$ & $15.55$ & $16.09$ & $15.79$ & $15.55$ \\ 
 \hline
 $\epsilon$ [m$^2$/s$^3$]  & $0.07 $ & $0.069$  & $0.1802$ & $0.07$ \\ 
 \hline
 $l_{int}$ [m] & $0.47$ & $0.488$ & $0.37$ & $0.47$\\
 \hline
 $k$ [m$^2$/s$^2$] &  $0.35$ & $0.37$ & $0.58$ & $0.35$\\
 \hline
 $T/\tau$ &  $743$ & $735$ & $1237$ & $744$ \\

\end{tabular}
\end{ruledtabular}
\end{table}

\begin{figure}[hb]
\begin{center}
\includegraphics[width=0.45\textwidth,trim={0cm 0cm 0cm 0cm},clip]{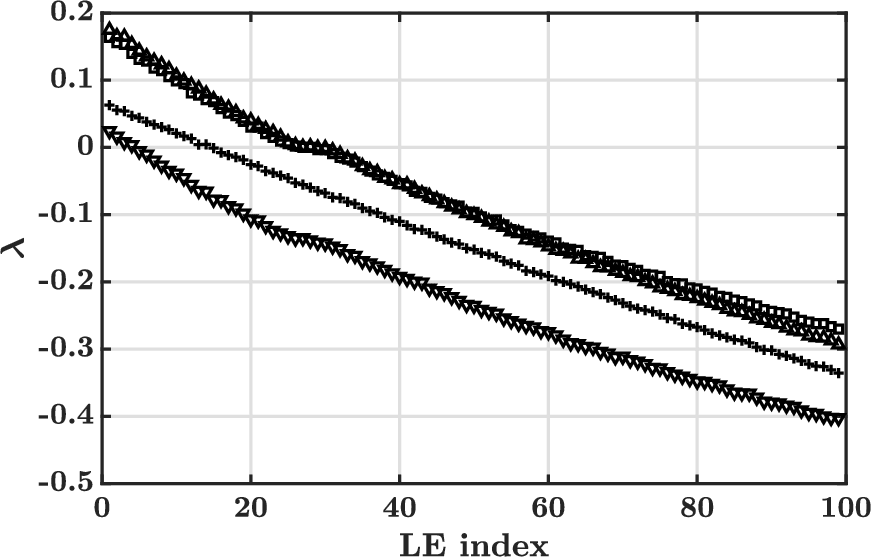}
\caption{First 100 LE for the linear forcing technique (Case 1, \mythicksquare{black}{white}), the linear forcing applied to the large scales only (Case 5, \mythicktriangle{black}{white}), the stochastic forcing technique (Case 6, \mythickcross{black}), the linear forcing techniques with the same force across simulations (Case 7, \mythickVtriangle{black}{white}).}
\label{fig:LE_forcing} 
\end{center}
\end{figure}

The first 100 Lyapunov exponents of all the forcing schemes for HIT are shown in Fig.~\ref{fig:LE_forcing}. Case 1 and 5 share similar values of the LE, implying similar chaotic behavior of the flow field. However, Case 6 shows significantly lower LE despite turbulent statistics that suggest a higher turbulence intensity. This feature can be understood by considering the last case. Cases 1 and 7 lead to the same statistics as they use the same forcing for the unperturbed simulation, but again, the LEs of Case 7 are significantly lower than Case 1. The level of chaoticity is therefore significantly reduced when all the realizations are forced in the same manner, which explains the results found in Case 6. This example shows that the Reynolds number alone is not enough to characterize the dynamics of a turbulent flow field and that the level of chaoticity is strongly dependent on the functional form of the forcing, and the dependence of the forcing on the perturbation. In the present case, it appears that a forcing insensitive to perturbations annihilate chaos in the flow field. More broadly, this study illustrates the potential of the Lyapunov analysis in comparing different models by assessing their effect on the dynamics of the flow field.

\section{Uncertainty estimation of attractor dimension and scaling}
\label{sec:app_dimuq}

Since the negative LEs are not available for the highest Reynolds number case (Case 4), the dimension needs to be estimated based on certain assumptions. From Fig.~\ref{fig:LEhighRE}, it can be reasonably inferred that the shape of the spectrum is similar across different Reynolds number. Further, it is approximated using a power-law form as $a(i-1)^{\alpha} + \lambda_1$, where $i$ is the LE index. The value of $a$ is chosen such that the fit passes through a certain $\lambda_i$. For the finest resolution, $i$ is chosen to be 49, which is the largest index available. For the other cases, $i$ is chosen such that $\lambda_i/\lambda_1$ is the same as for the highest $Re_{\lambda}$. As a result, the fit is parameterized using only one variable: $\alpha$. 

\begin{figure}[hb]
\begin{center}
\includegraphics[width=0.4\textwidth,trim={0cm 0cm 0cm 0cm},clip]{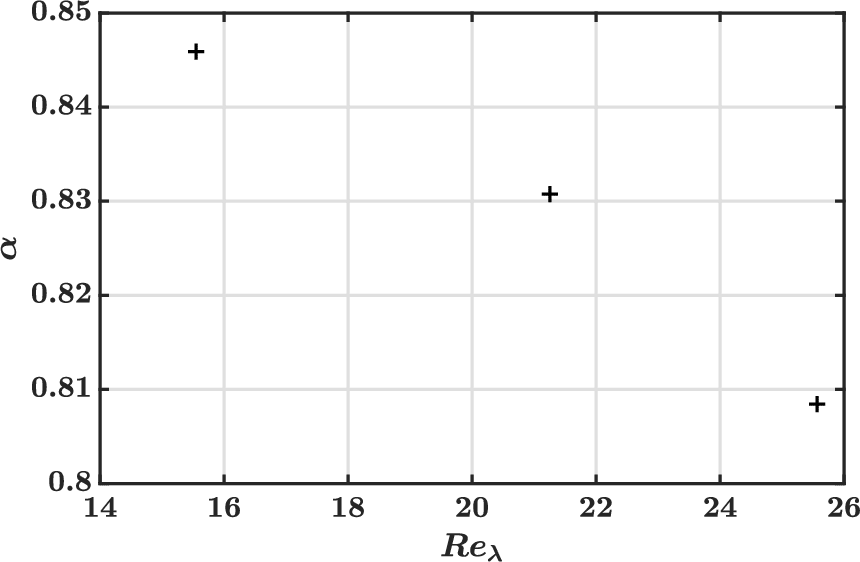}
\caption{Value of $\alpha$ obtained from the fit applied to the LE computed with Case 1, 2 and 3.}
\label{fig:alpha}
\end{center}
\end{figure}

In Fig.~\ref{fig:alpha}, the parameter $\alpha$ is plotted for the three lowest $Re_{\lambda}$ cases. Based on this result, $\alpha$ is extrapolated linearly with respect to the $Re_{\lambda}$.  For this purpose, the slope is estimated using any two combinations of the three available points, and the extrapolation is done starting from any of the three available points. Nine possible values for $\alpha$ are obtained. All of these values are used to fit the first 49 LE of largest $Re_{\lambda}$ case. Finally, the average of all these possible dimensions is taken to be the best dimension estimate. The uncertainty estimate is obtained from the maximal and the minimal dimensions obtained using all the possible $\alpha$.

At this stage, an estimate of the attractor dimension for Cases 1, 2, 3 and 4 along with an uncertainty estimate has been obtained. In addition, an uncertainty estimate for the mean turbulent quantities (here, the Kolmogorov length scale $\eta$) can be obtained using the method outlined in Ref.~\cite{uncertaintyDNS_oliver}. Given these dimension estimates for the four Reynolds numbers, the goal is to determine the scaling of the dimension as a function of $\frac{L}{\eta}$, where $L=2\pi$. In other terms, one wants to fit a curve of the form $ax^{b}$ through the data points, get an estimate as well as an uncertainty estimate for $b$. To do so, an approach similar to the one used in Appendix~\ref{sec:app_dimuq} is used. Different possible scalings are generated and are used to obtain an uncertainty estimate. 

\begin{figure}[hb]
\begin{center}
\includegraphics[width=0.4\textwidth,trim={0cm 0cm 0cm 0cm},clip]{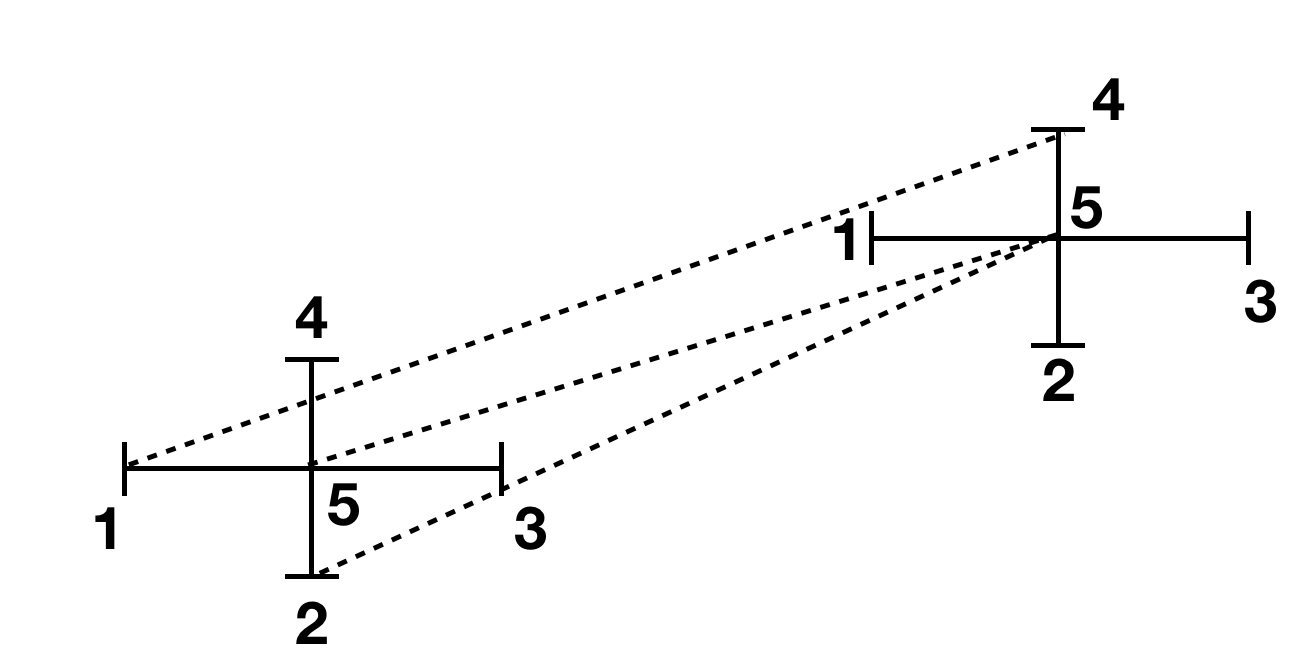}
\caption{Schematic of the uncertainty estimation method for the attractor dimension scaling. For each case, 5 points are considered using the uncertainty estimate of the dimension. The horizontal error bar represents the statistical error for the value of the average $\frac{L}{\eta}$, the vertical error bar contains the statistical uncertainty and the extrapolation uncertainty used to compute the dimension. The dashed lines represent examples of the fits considered.}
\label{fig:uqscaling}
\end{center}
\end{figure}

Schematically, the points that are used to compute the scaling are arranged in a manner illustrated by Fig.~\ref{fig:uqscaling}. The x-axis represents the $\frac{L}{\eta}$ and the y-axis represent the estimated dimension. For each Cases 1-4, an uncertainty estimate for both axes is available. In Fig.~\ref{fig:uqscaling}, the data of only two cases (for simplification) through which a fit goes are schematically shown. Given the estimates, many different fits are possible. For each Case, one can consider that one fit intersect any of the five points indicated in Fig.~\ref{fig:uqscaling} (represented by the dashed lines), which implies that there are then $5^4$ possible fits for the data.

All of these fits are generated here, resulting in $5^4$ possible values for $b$. 
At each data point, if a fit uses any point different than the point 5 shown in Fig.~\ref{fig:uqscaling}, it is attributed a lower weight equal to the ratio $\frac{e^{-1/2}}{e^0}$. The lowest possible weight for a fit is then  $\frac{e^{-2}}{e^0}  \approx 0.135$. The estimate for $b$ and for its uncertainty estimate are obtained as a weighted average of the $b$ and $b^2$ values generated by each fit. In the end, this procedure leads to the scaling $b=2.8 \pm 0.095$.

The same procedure can be used for the three lowest Reynolds numbers in order to remove the possible influence of extrapolation errors. In that case, $b=2.35 \pm 0.055$. The discrepancy between both values of $b$ could indicate that the dimension scales differently with $L/\eta$ at different Reynolds numbers. This property was also observed for two-dimensional turbulence \cite{kevlahan2007scaling}.

\section{Consequences of the energy spectrum of the GSV on the energy spectrum of the CLV}
\label{sec:app_clv}

From Fig.~\ref{fig:spectrum}, the energy spectra of the GSVs do not change much based on the index. This feature is used to relate the energy spectra of the GSV to that of the CLV. In particular, it is shown that the first few CLV cannot be localized in Fourier space. Let $\bphi_i$ be the i-th CLV and $\bgsv_i$ be the i-th GSV, then the CLV can be expressed as \cite{ginelli2013covariant}:
\begin{equation}
    \forall \bx ,~~\bphi_j(\bx) = \sum_{i=1}^j a_{i,j} \bgsv_i(\bx), 
    \label{eq:clv2}
\end{equation}
where $\bx$ denotes the physical space location and $a_{i,j} \in \mathbb{R}$. For ease of notation, the physical space location $\bx$ is dropped from the notation. Since $\bphi_j$ are normalized, 
\begin{equation}
    \sum_{i=1}^j a_{i,j}^2 = 1.
    \label{eq:norm}
\end{equation}

Equation.~\ref{eq:clv2} can be rewritten using the Galerkin projection onto the Fourier modes as
\begin{equation}
    \forall \bk,~~\widehat{\bphi}_j(\bk) = \sum_{i=1}^j a_i \widehat{\bgsv}_i(\bk),
    \label{eq:galerkinclv}
\end{equation}
where $\kappa$ is a three dimensional wavenumber, $\widehat{\bphi}_j(\bk)$ and $\widehat{\bgsv}_i(\bk) \in \mathbb{C}^3$ are the Fourier amplitudes of the mode $\bk$.

Let $K_0 \in \mathbb{R}$. The goal is to find a relation between the energy at the wavenumber $K_0$ for the j-th CLV, which is defined as
\begin{equation}
 E_{\bphi_j}(K_0) = \mathop{\sum_{\bk}}_{|\bk|=K_0} \widehat{\bphi_j}(\bk)^* \cdot \widehat{\bphi_j}(\bk),
\end{equation}
where $\cdot^*$ denotes the complex conjugate. Using Eq.~\ref{eq:galerkinclv},
\begin{multline}
 E_{\bphi_j}(K_0) = \sum_{i=1}^j a_{i,j}^2 E_{\bgsv_i}(K_0) + \\ \mathop{\sum_{\bk}}_{|\bk|=K_0} \mathop{\sum_{k,l \leq j}}_{k \neq l} a_{k,j} a_{l,j} (\widehat{\bgsv}_k(\bk)^* \cdot \widehat{\bgsv}_l(\bk) + \widehat{\bgsv}_l(\bk)^* \cdot \widehat{\bgsv}_k(\bk)). 
\end{multline}
Due the fact that the energy spectra of the GSV are relatively independent of the index, Eq.~\ref{eq:norm} can be used to write
\begin{equation}
 E_{\bphi_j}(K_0) \approx E_{GSV}(K_0) + 2 \sum_{k,l,~k\neq l}a_{k,j} a_{l,j} E_{corr,kl}(K_0),
 \label{eq:approx1}
\end{equation}
where  $E_{GSV}(K_0)$ is the energy of the GSV for Fourier modes of amplitude $K_0$, and $ E_{corr,kl} (K_0) = \frac{1}{2} \mathop{\sum_{\bk}}_{|\bk|=K_0} (\widehat{\bgsv}_k(\bk)^* \cdot \widehat{\bgsv}_l(\bk) + \widehat{\bgsv}_l(\bk)^* \cdot \widehat{\bgsv}_k(\bk)) $

\begin{figure}[hb]
\begin{center}
\includegraphics[width=0.4\textwidth]{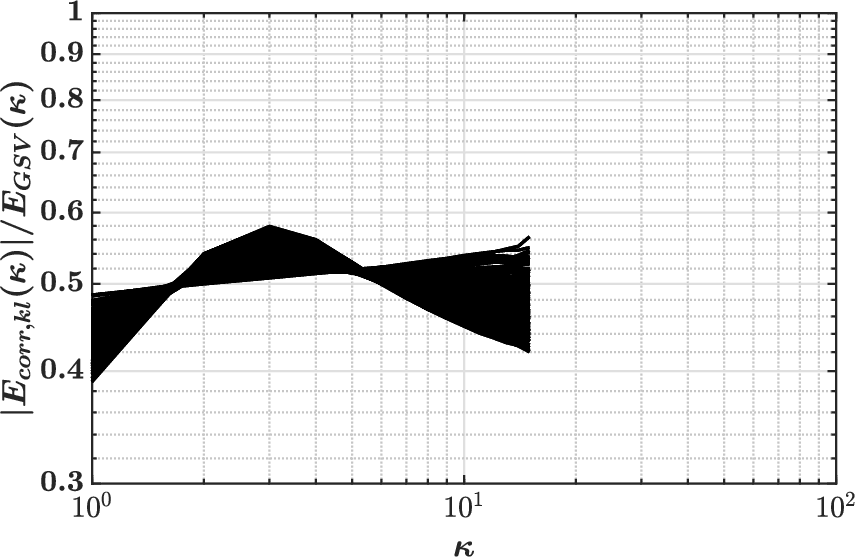}
\caption{Ratio $\frac{|E_{corr,kl}(\bk)|}{E_{GSV}(\bk)}$ plotted for Case 1 and the first 19 pairs of LV, plotted against the wavenumber amplitude.}
\label{fig:corr}
\end{center}
\end{figure}

The second term of the right-hand side of Eq.~\ref{eq:approx1} is what distinguishes the energy spectrum of the GSV from the spectrum of the CLV. Note that this term is real but not necessarily positive. This term can be estimated from the database generated in the present study. In Fig.~\ref{fig:corr}, the ratio  $\frac{|E_{corr,kl}(\bk)|}{E_{GSV}(\bk)}$ is plotted against $\bk$ for all combinations of GSVs from 1 through 19 indices. For all of these combinations, the ratio can be reasonably estimated to be equal to $0.5$ for all the wavenumber amplitudes.

Further, to bound the CLV spectrum, bounds on the product of coefficients $a_{k,j} a_{l,j}$ are needed. To do so, it can be recognized that the possible $a_{p,j}$ are the points located on the j-sphere (hypersphere of dimension j) of radius 1 which leads to 

\begin{equation}
    \forall p,q,~ |a_{p,j} a_{q,j}| = | cos \gamma_p cos \gamma_q \prod_{k=1}^{p-1} sin \gamma_k \prod_{l=1}^{q-1} sin \gamma_l |,
\end{equation}
where $\gamma_i \in [0, \pi]$ if $i \leq j-2$ and $\gamma_{j-1} \in [0, 2\pi]$. By recognizing that $|a_{p,j} a_{q,j}|$ either contains a product of the form $|cos \gamma_p sin \gamma_p|$ or $|cos \gamma_q sin \gamma_q|$,
\begin{equation}
    \forall p,q,~ |a_{p,j} a_{q,j}| \leq 0.5
\end{equation}

\begin{figure}[hb]
\begin{center}
\includegraphics[width=0.4\textwidth]{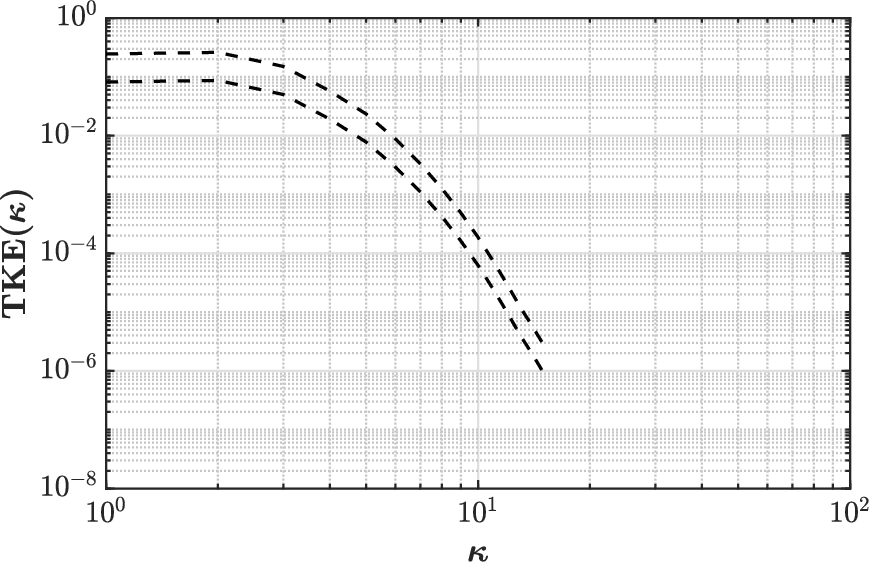}
\caption{Estimated bounds for the energy spectrum of the second CLV.}
\label{fig:bounds}
\end{center}
\end{figure}

Using the triangle inequality, one can then obtain bounds for the energy spectrum of $\bphi_j$:
\begin{multline}
    max(E_{GSV}(K_0) - \frac{1}{2} \binom{j}{2} E_{GSV}(K_0), 0) \lesssim \\ E_{\bphi_j}(K_0) \lesssim \\ E_{GSV}(K_0) + \frac{1}{2} \binom{j}{2} E_{GSV}(K_0).
\end{multline}
Note that the bounds increase in range as the index of the CLV increases. As a result, these bounds are useful only for the first few CLVs. For example, the bounds for the second CLV can be obtained as
$0.5 E_{GSV}(K_0) \lesssim E_{\bphi_2}(K_0) \lesssim 1.5 E_{GSV}(K_0)$. These bounds are shown in Fig.~\ref{fig:bounds}, and follow the shape of the GSV spectrum. As a result, the first few CLV are not localized in Fourier space.

\begin{acknowledgments}
This work was financially supported by an AFOSR research grant (FA9550-15-1-0378) with Chiping Li as program manager. The authors thank NASA HECC for generous allocation of computing time on NASA Pleiades machine. Useful discussions with Robert D. Moser, Guillaume Blanquart and Masanobu Inubushi are gratefully acknowledged.
\end{acknowledgments}

\bibliography{master}

\end{document}